\documentclass[lettersize,journal]{IEEEtran}\IEEEoverridecommandlockouts
\usepackage[utf8]{inputenc}

\usepackage{derivative}
\usepackage{xspace}
\usepackage{url}
\usepackage{bbm}
\usepackage{graphicx}
\usepackage{epstopdf}
\usepackage{paralist}
\usepackage[normalem]{ulem}
\usepackage{fancyref} 

\usepackage{amsmath}
\usepackage{amsfonts}
\usepackage{amssymb}

\usepackage{bm}

\usepackage{algorithm}
\usepackage{algpseudocode}

\usepackage{textcomp}
\usepackage{array}
\usepackage{arydshln}
\usepackage{mwe}
\usepackage{url}
\usepackage{siunitx} 
\usepackage{tikz}
\usepackage{gensymb}

\usepackage[inline]{enumitem}
\usepackage{xcolor}
\definecolor{links}{rgb}{0.3,0,0}   
\definecolor{urls}{rgb}{0,0,0.8}    
\definecolor{cites}{rgb}{0,0,0.6}   
\usepackage[colorlinks,hyperindex,linkcolor=links,citecolor=cites,urlcolor=urls]{hyperref} 

\definecolor{silver}{cmyk}{0,0,0,0.3}
\definecolor{navy}{cmyk}{0.8,0.5,0,0}
\definecolor{lightblue}{cmyk}{0.35,0.11,0,0}
\definecolor{orange}{cmyk}{0,0.57,0.86,0}
\definecolor{yellow}{cmyk}{0,0,0.9,0.0}
\definecolor{reddishyellow}{cmyk}{0,0.22,1.0,0.0}
\definecolor{lightred}{cmyk}{0,0.820,0.753,0.0}
\definecolor{black}{cmyk}{0,0,0.0,1.0}
\definecolor{white}{cmyk}{0,0,0.0,0}
\definecolor{purple}{cmyk}{0.64,0.83,0,0}
\definecolor{darkYellow}{cmyk}{0,0,1.0,0.5}
\definecolor{darkSilver}{cmyk}{0,0,0,0.1}
\definecolor{lightyellow}{cmyk}{0,0,0.3,0.0}
\definecolor{lighteryellow}{cmyk}{0,0,0.1,0.0}
\definecolor{lightestyellow}{cmyk}{0,0,0.05,0.0}
\definecolor{darkblue}{cmyk}{0.98,0.89,0,0.11}
\definecolor{bluel1}{cmyk}{0.5,0.05,0.05,0.05}
\definecolor{darkred}{cmyk}{0,0.89,0.7,0.55}
\definecolor{magenta}{rgb}{1.0, 0.0, 1.0}
\definecolor{cyan}{rgb}{0.0, 1.0, 1.0}
\usepackage{dsfont}
\usepackage{footnote}
\usepackage{balance}
\usepackage{float}
\usepackage{cite}
\newcommand{\infden}{\imath_{s}}

\usepackage{wasysym}
\usepackage{amssymb}
\usepackage{tabularx,booktabs}
\newcolumntype{L}[1]{>{\raggedright\let\newline\\\arraybackslash\hspace{0pt}}m{#1}}
\newcolumntype{C}[1]{>{\centering\let\newline\\\arraybackslash\hspace{0pt}}m{#1}}
\newcolumntype{R}[1]{>{\raggedleft\let\newline\\\arraybackslash\hspace{0pt}}m{#1}}
\usepackage{boldline}

\usepackage{subcaption}
\usepackage{pgfplots}
\usepgfplotslibrary{groupplots}
\usetikzlibrary{pgfplots.groupplots}
\usepackage{bm}
\usepackage{vmr-symbols-vecbold}
\usepackage{standard-macros}

\newcommand{\rate}{\mathrm{R}}
\newcommand{\mN}{\mathrm{N}}
\newcommand{\mM}{\mathrm{M}}
\newcommand{\transpose}{\mathrm{T}}

\safemath{\rmatW}{\mathbb{W}}
\safemath{\rmatY}{\mathbb{Y}}
\safemath{\rmatZ}{\mathbb{Z}}
\safemath{\matR}{\boldsymbol{\mathsf{R}}}
\safemath{\matZ}{\boldsymbol{\mathsf{Z}}}
\safemath{\matQ}{\boldsymbol{\mathsf{Q}}}
\safemath{\matY}{\boldsymbol{\mathsf{Y}}}
\safemath{\matP}{\boldsymbol{\mathsf{P}}}
\safemath{\matV}{\boldsymbol{\mathsf{V}}}
\safemath{\matI}{\boldsymbol{\mathsf{I}}}
\safemath{\matJ}{\boldsymbol{\mathsf{J}}}

\newcommand{\nb}{n_{\text{b}}}
\newcommand{\ns}{n_\text{{s}}}
\newcommand{\nc}{n_{\text{c}} }
\newcommand{\np}{n_{\text{p}} }
\newcommand{\tp}{t_{\text{p}} }
\newcommand{\ts}{t_{\text{s}} }
\newcommand{\spk}{_{k,\ell}}
\newcommand{\spkk}{_{k-1,\ell}}

\newcommand{\hatVech}{\skew{-5}\hat{\vech}}

\safemath{\mfn}{\mathfrak{N}_{\zeta}}
\DeclareMathOperator{\re}{\mathrm{Re}}
\DeclareMathOperator{\im}{\mathrm{Im}}

\algrenewcommand\algorithmicrequire{\textbf{Input:}}
\algrenewcommand\algorithmicensure{\textbf{Provide:}}

\begin{document}
\title{Is Synchronization a Bottleneck for Pilot-Assisted URLLC Links?}
\author{\IEEEauthorblockN{A. Oguz Kislal, Madhavi Rajiv, Giuseppe Durisi,~\IEEEmembership{Senior Member,~IEEE}, Erik G. Ström,~\IEEEmembership{Fellow,~IEEE}, Urbashi Mitra,~\IEEEmembership{Fellow,~IEEE}}
\thanks{A. Oguz Kislal, Giuseppe Durisi, and Erik. G. Str\"om are with the Department of
Electrical Engineering, Chalmers University of Technology, Gothenburg 41296,
Sweden (e-mail: \{kislal,durisi,erik.strom\}@chalmers.se). 
Madhavi Rajiv and Urbashi Mitra are with the Ming Hsieh Department of Electrical Engineering, University of Southern California, Los Angeles CA 90089, USA (email: \{rajiv,ubli\}@usc.edu). This work has been funded in part by the following grants: Swedish Research Council 2018-04359, NSF CCF-1817200, DOE DE-SC0021417, NSF CCF-2008927, NSF CCF-2200221, ONR 503400-78050, ONR N00014-15-1-2550, NSF A22-2666-S003, ONR N00014-22-1-2363, NSF KR705319.
 
}
\thanks{Parts of this work have been presented at the International Conference on Communication, (ICC), Denver CO, U.S.A., Jun. 2024~\cite{Kislal2024_ICCPaper}.}
}

\maketitle
\sloppy
\begin{abstract}
We propose a framework to evaluate the so-called random-coding union bound with parameter $s$ (RCUs) on the achievable error probability in the finite-blocklength regime for a pilot-assisted transmission scheme operating over an imperfectly synchronized and memoryless block-fading waveform channel. Unlike previous results, which disregard the effects of imperfect synchronization, our framework utilizes pilots for both synchronization and channel estimation. Specifically, we provide an algorithm to perform joint synchronization and channel estimation, and verify its accuracy by observing its tightness in comparison with the Cramer-Rao bound. Then, we develop an RCUs bound on the error probability, which applies for a receiver that treats the estimates provided by the algorithm as accurate. Additionally, we utilize the saddlepoint approximation to provide a numerically efficient method for evaluating the RCUs bound in this scenario. 
Our numerical experiments verify the accuracy of the proposed approximation. Moreover, when the delays are modeled as fully dependent across fading blocks, numerical results indicate that the number of pilot symbols needed to estimate the fading channel gains to the level of accuracy required in ultra-reliable low-latency communication is also sufficient to acquire sufficiently good synchronization. However, when the delays are modeled as independent across blocks, synchronization becomes the bottleneck for the system performance.


\end{abstract}


\section{Introduction}
\label{sec:Introduction}
Advancements in technology have allowed for the rise of applications like remote surgery \cite{Kolovou2021, Liou2020}, factory automation \cite{Peng2023}, and autonomous driving that significantly impact various aspects of people's lives. These types of applications create substantial data traffic that future cellular communication technologies will need to accommodate. 5G has the goal to support the diverse requirements of these technologies.

The scenario considered in this paper, denoted as \textit{ultra-reliable low-latency communication} (URLLC), is designed for mission-critical applications targeting $99.999 \% $ reliability with end-to-end latency as low as 1 ms \cite{3GPP_URLLC}. For 6G, the targets will be even more aggressive: $10^{-5}$ -- $10^{-7}$ packet error rate with a more stringent latency constraint \cite{Tataria2021, Chentao2023,Coskun2019}.   

A key feature of URLLC traffic is the frequent use of small information payloads, which are transmitted in short packets consisting of a limited number of encoded symbols representing the payload data. In URLLC, there are limitations on the signal duration and available bandwidth, due to latency requirements and the need to orthogonalize multiple-user transmissions to mitigate multi-user interference. As a consequence, the conventional asymptotic performance metrics commonly employed in the design of communication systems, namely the ergodic and outage rates, are not suitable for the short-packet regime \cite{durisi16-09a}. Because of its relevance for URLLC, the field of finite-blocklength information theory has been studied extensively, particularly following the seminal work \cite{Polyanskiy2010}, which offers a precise understanding of the tradeoff between error probability and packet size, for a given SNR and transmission rate, when operating with finite blocklengths.

In this paper, we focus on communication over memoryless block-fading channels with imperfect synchronization,\footnote{
Throughout this paper, by synchronization, we mean time synchronization, or, more specifically, the estimation of the integer and fractional delays of the samples at the output of the upsampled receiver matched filter.} with the goal of understanding the impact of synchronization accuracy on the overall system performance. Benchmarking URLLC systems with such goals often requires the use of approximation techniques due to the computational expense of evaluating finite blocklength bounds exactly. Such approximation techniques are critical to enable the use of these bounds within URLLC optimization routines such as resource-allocation \cite{Anand2018,Ghanem2022} and scheduling algorithms \cite{Darabi2022, Chen2023}.

The computationally efficient approximation proposed in this paper is based on the saddlepoint method \cite{jensen95-a}, which is used to evaluate the random coding union bound with parameter $s$ (RCUs) proposed in~\cite{Martinez2011}. This bound is well-suited for communication over fading channels, as it can be applied to both the optimal non-coherent maximum-likelihood (ML) decoder and more practically significant decoders that utilize pilot-assisted transmission (PAT), see, e.g., \cite{Ostman2019}. Specifically, the decoder considered in this paper is the so-called scaled nearest-neighbor (SNN) decoder \cite{Lapidoth2002}, which minimizes the error probability when synchronization and channel estimation are perfect, but is suboptimal otherwise. The evaluation of the RCUs bound for URLLC scenarios can be computationally expensive, and, for some applications, such bounds need to be approximated by methods similar to the one proposed in \cite{Kislal2023}. 

Most existing approaches for benchmarking URLLC systems disregard synchronization errors. In these approaches, an upper bound on the error probability is obtained by evaluating the tail probability of a sum of independent random variables. This allows for approximations such as the \textit{normal approximation} \cite{Polyanskiy2010}, or the more accurate \textit{saddlepoint approximation} \cite{Feller1971,jensen95-a}. The saddlepoint approximation has been applied to the case of an optimal ML decoder in \cite{Lancho2020}, of PAT transmission and SNN decoding over single-input single-output (SISO) and multiple-input multiple-output (MIMO) channels in \cite{Ferrante2018}, and of PAT transmission and SNN decoding over massive MIMO for both single and multiple fading blocks in \cite{Ostman2019,ostman2020,Kislal2023}. 
In \cite{Yury2013}, the problem of joint synchronization and decoding is analyzed from a finite-blocklength perspective. There, it is shown that one can achieve performance close to the perfectly synchronized case even for an exponentially large asynchronism level. However, asynchronism is modeled only at the symbol level, and imperfect synchronization results only in an incorrect estimate of the location of the transmitted codeword. 

In this paper, we consider the more practically relevant case in which asynchronism is at the waveform channel, and imperfect synchronization yields intersymbol interference after matched filtering and sampling. As we shall see, this implies that the random variables that need to be analyzed to obtain upper bounds on the error probability are, in general, dependent, and the saddlepoint approximations provided in \cite{Ostman2019,ostman2020,Ferrante2018,Kislal2023} are therefore not applicable. However, under certain assumptions on the dependence between the random variables, saddlepoint approximations can still be derived as we do herein.




\paragraph*{Contributions}
We consider the problem of transmitting short packets over a SISO memoryless block-fading waveform channel with unknown delays, and analyze the impact of imperfect synchronization and channel estimation on system performance in the URLLC regime. 
Specifically, we focus on the extreme cases of delays being independent across all fading blocks, or delays being fully dependent across fading blocks, and design a synchronization and channel estimation algorithm that can be used in both scenarios.
We verify the accuracy of the proposed algorithms by presenting a comparison with the Cramer-Rao bound (CRB). We then derive an RCUs bound for the case of PAT and SNN decoding to evaluate the error probability achievable by a decoder that treats the estimates for the delay and the channel gains provided by our algorithm as perfect. Finally, we provide a novel saddlepoint approximation for this bound based on \cite[Ch.~6]{jensen95-a}. The novelty of the proposed approximation is that it accounts for the specific dependence between random variables that arises due to imperfect synchronization. Finally, we provide numerical results to illustrate the accuracy of the proposed approximation and its utility in obtaining insights into the design of URLLC links. 

In this paper, we extend the results in~\cite{Kislal2024_ICCPaper} by providing a complete derivation of the upsampled output of the matched filter at the receiver. We also explain how to evaluate in closed form the moment-generating function appearing in the saddlepoint approximation and provide a concrete example for the case of BPSK transmission. Finally, with extensive numerical experiments, we clarify, among other aspects, the optimal number of pilot symbols to be used for synchronization and channel estimation.



\paragraph*{Notation} 
We denote random vectors and random scalars by upper-case boldface letters such as
$\rvecx$ and upper-case standard letters,
such as $\rndx$, respectively. 
Their realizations are indicated by lower-case letters of the same font. We use upper-case letters of two special fonts to denote deterministic matrices (e.g., $\matY$), and use the notation $[\matY]_{ij}$ to refer to the element on the $i$th row and 
$j$th column of the matrix $\matY$.
To avoid ambiguities, we use another font, such as $\rate$ for rate, to denote constants that are typically
capitalized in the literature. The circularly symmetric Gaussian distribution is denoted by $\jpg(\mu,
\sigma^2)$, where $\mu$ and $\sigma^2$ denote the mean and the variance, respectively. The superscripts $(\cdot)^T$ and $(\cdot)^H$ denote transposition and Hermitian transposition, respectively. 
We write $\log(\cdot)$  to denote the natural logarithm,  
$\vecnorm{\cdot}$ stands for the $\ell_2$-norm, $\Prob[\cdot]$
for the probability of an event, $\Exop[\cdot]$ for the expectation operator, $*$ for the convolution operation, $Q(\cdot)$ for the Gaussian $Q$-function, $\ind{\cdot}$ for the indicator function,
$\vecone_{u}$ for the all-one column vector with a length $u$, and $\boldsymbol{\mathsf{I}}_m$ for the $m\times m$ identity matrix, respectively. Finally, for two functions $f(n)$ and $g(n)$, the notation $f(n) = o(g(n))$ means that $\lim_{n\to\infty} f(n)/g(n)=0$ and the notation $f(n) =\mathcal{O}(g(n))$ means that $\limsup_{n\to\infty} \abs{f(n)/g(n)}<\infty$.
 
\section{System Model}
\label{sec:SystemModel}
\subsection{Overview}
\begin{figure}
    \centering
    \includegraphics[width=1\columnwidth]{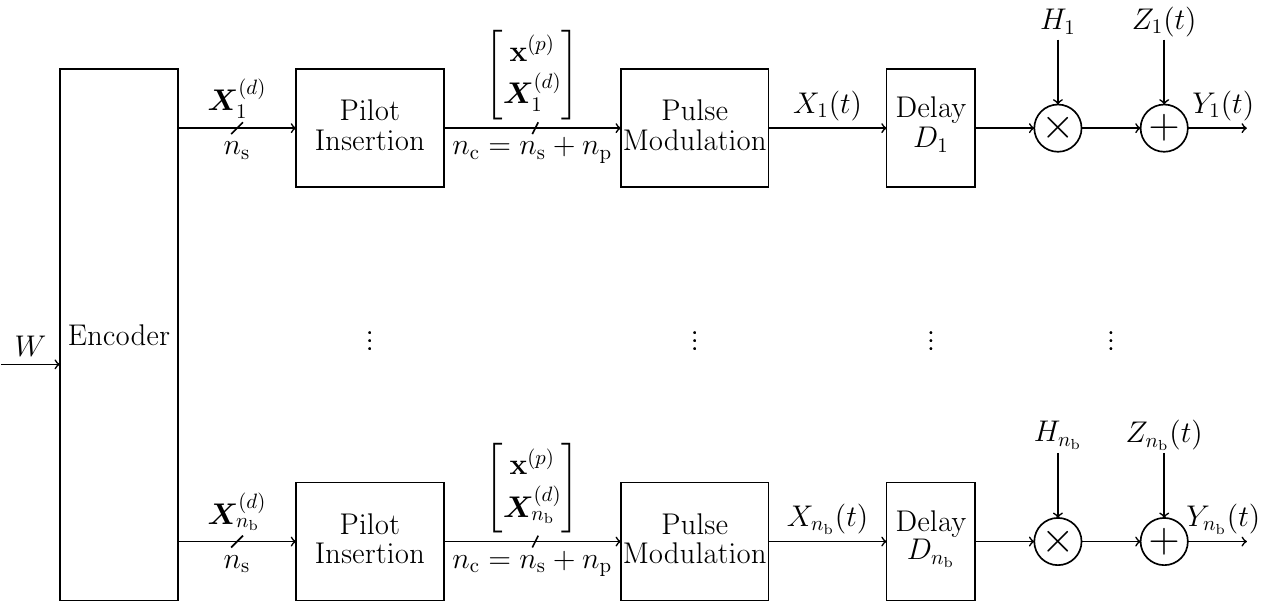}
    \caption{Block diagram of the system model. In the case of fully dependent delays}, $\rndd_1 = \ldots = \rndd_{\nb}$.
    \label{fig:block}
\end{figure}

We consider pilot-assisted transmission of a uniformly distributed message over a SISO block-flat-fading channel with unknown delay. We assume that the channel stays constant over the transmission of a block of $\nc$ channel uses and that it changes independently across blocks consisting of non-overlapping channel uses. Independence may be achieved by spacing the blocks sufficiently in frequency or time. 
Each transmitted packet spans $\nb$ such fading blocks.  The setup is illustrated by the block diagram in Fig.~\ref{fig:block}, in which each branch corresponds to the transmission over one fading block.
The encoder maps the realization of the random message $\rndw$ to a complex-valued codeword of length $\nb\ns$, which, in turn, is split into $\nb$ subcodewords $\{\vecx_{\ell}^{(d)} \}_{\ell=1}^{\nb}$ of length $\ns$ 
\begin{align}
    \label{eq:rvec:d:def}
  \vecx^{(d)}_{\ell} = 
  \begin{bmatrix}
   x_{1,\ell}^{(d)} & \cdots & x_{\ns,\ell}^{(d)}   
  \end{bmatrix}^\transpose\in\mathbb{C}^{\ns},
\end{align}
where $\ell = 1,\ldots,\nb$ and $\nb$ is the number of fading blocks used for the transmission of the message. A pilot sequence  
\begin{align}
    \label{eq:rvec:p:def}
  \vecx^{(p)} = 
  \begin{bmatrix}
   x_{1}^{(p)}& \ldots & x_{\np}^{(p)}   
  \end{bmatrix}^\transpose\in\mathbb{C}^{\np}
\end{align}
is prepended to each codeword, to form $\nb$ subpackets of length $\nc = \np+\ns$, which are transmitted over different channel fading blocks and experience potentially different propagation delays. We assume that $\rndw$ is drawn uniformly from the set $\{1,2,\ldots, \lceil \exp(\nb\nc\rate) \rceil \}$ and that $\lceil \exp(\nb\nc\rate) \rceil$ is the number of codewords in the codebook. As a consequence, $\rate$ is the transmission rate in nats per symbol.

We assume that the data and pilot symbols are subject to the power constraint $\Exop[\| \rvecx^{(d)}_{\ell}\|^2] = \ns\rho$ and  $\|\vecx^{(p)}\|^2 = \np\rho$, respectively.

In the following subsections, we will define the pulse shaping, introduce the propagation delays in the waveform domain, and describe the processing at the receiver considered in this paper.   

\subsection{Signal Model}\label{subsec:signal-model}
The $\np$ pilot symbols within each subpacket are used to form the continuous-time pilot signal $x^{(p)}(t)$ defined as 
\begin{equation} \label{eqn:cts-pilot}
x^{(p)}(t) = \sum_{k=1}^{\np} x_k^{(p)} s_{\tp}(t-(k-1)\tp),
\end{equation}
where $s_{\tp}(t)$ is a rectangular pulse\footnote{We consider rectangular pulses for simplicity. Our analysis can be generalized to other pulse shapes by using the corresponding matched filter in~\eqref{eq:matched-filter}.} with normalized energy and support of size $\tp$, i.e.,
\begin{eqnarray}
s_{\tp}(t) =    \begin{cases}
        \frac{1}{\sqrt{\tp}}, &  t \in [0,\tp)\\
        0, & \text{otherwise }.
    \end{cases}
\end{eqnarray}

The data symbols in the $\ell$th subpacket are sent after the pilot symbols via the continuous-time signal
\begin{align} 
\label{eqn:cts-data}
x_{\ell}^{(d)}(t) &= \sum_{k=1}^{n_s} x^{(d)}_{k,\ell} s_{\tp}(t-(k-1) \tp - n_pt_p).
\end{align}
The total continuous-time signal corresponding to the $\ell$th subpacket is put through a flat-fading channel to obtain the received continuous signal 
\begin{align}\label{eq:received-continuous}
    Y_{\ell}(t) &= H_{\ell}\ltrp{x^{(p)}(t-D_{\ell}) +  x^{(d)}_{\ell}(t-D_{\ell}) }  + Z_{\ell}(t),
\end{align}
where $\rndh_{\ell}$ denotes the scalar random fading complex channel gain for the $\ell$th fading block, $\rndd_{\ell}$
is the time delay for the $\ell$th fading block, which we assume to be uniform in $[0, d_{\text{max}}]$,
and $\rndz_{1}(t), \ldots, \rndz_{\nb}(t)$ are independent white complex Gaussian processes with power spectral density $N_0$. For simplicity, we set $N_0 = 1$.

%
%

We highlight that other types of asynchronism beyond the one modeled in~\eqref{eq:received-continuous}, including clock and
		      carrier asynchronism may occur in a wireless communication system.
		      These other forms of asynchronism are not captured directly by~\eqref{eq:received-continuous} (although
		      impairments such as carrier-phase offset can be modeled by modifying the statistics
		      of the fading coefficients $H_{\ell}$), and their impact on performance will not be
		      discussed in the present paper.

Depending on the type of diversity being used, the  delays $D_\ell$ in~\eqref{eq:received-continuous} may be dependent across $\ell$.  
For simplicity, we focus on the two extreme cases 
of delays being independent across fading blocks, where $D_1, \ldots, D_{\nb}$ are \iid and the case of delays being fully dependent across fading blocks, where $D_1 =~ \cdots =~ D_{\nb} =~ D$. 
In the former case, information about the pair $(D_\ell, H_\ell)$ is found only in $Y_\ell(t)$, and the estimation problem decouples into $\nb$ separate problems, one per block. However, in the case of fully dependent delays, it will be advantageous to jointly estimate $(D, H_1, \ldots, H_{\nb})$. We next introduce synchronization and channel estimation algorithms for both cases.

\subsection{Synchronization and Channel Estimation Phase}\label{subsec:synch}
The receiver uses the knowledge of the pilot sequence to estimate the propagation delays and the fading-block gains.
Throughout, we assume that the receiver employs synchronization and channel estimation algorithms that take an upsampled version of the received signal as the input. For a given upsampling rate $\mN$, let $\ts$ be the sampling interval; then, $\tp = \ts \mN$ is the period of the pulses used to construct the continuous time pilot and data signals.
We also let $x_{\mN,n}^{(p)} = x_{\ceil{n/\mN}}^{(p)}$ be the $n$th element of the upsampled vector of pilot symbols, where $n = 1,\ldots,\mN \np$. 
In order to obtain an upsampled received signal, we process and sample the received signal $\rndy_{\ell}(t)$ as 
\begin{equation} \label{eqn:synch_sample}
   \rndy^{(p)}_{m,\ell} = \left(Y_{\ell}*s_{\ts}\right)\left(m\ts\right),
\end{equation}
where $m = 1, \ldots, \mM$, and the sampling endpoint is chosen to capture all pilot symbols as
\begin{equation}
\label{eq:Mdef}
\mM = \lceil d_{\text{max}}/{\ts} \rceil + n_p \mN.    
\end{equation}

Let $\rvecy_{\ell}^{(p)} = [\rndy_{1,\ell}^{(p)}, \ldots, \rndy_{\mM,\ell}^{(p)}]^T$. To express $\rvecy_{\ell}^{(p)}$ compactly, we shall first isolate the contribution to $\rvecy_{\ell}^{(p)}$ of the pilot signal $x^{(p)}(t)$ in \eqref{eqn:cts-pilot}. In view of \eqref{eqn:synch_sample}, we start by noting that the convolution between $x^{(p)}(t)$ and $s_{\ts}(t)$ results in the signal
\begin{equation}\label{eq:matched-filter}
    (x^{(p)} * s_{\ts})(t) =  \sum_{n=1}^{\mN n_p} x_{\mN, n}^{(p)} r_{\ts}(t- (n-1)\ts),
\end{equation}
where $r_{\ts}(t)$ is a triangular pulse with duration $2\ts$ and unit peak amplitude
\begin{align}
    r_{\ts}(t) = (s_{\ts}*s_{\ts}) (t)
    =
    \begin{cases}
        t/\ts, & 0\le t< \ts\\
        2-t/\ts, & \ts\le t\le 2\ts\\
        0, & \text{otherwise}.
    \end{cases}
\end{align}
Thus, for any $e \in [0,\ts)$
\begin{equation} 
    \label{eq:pilot:ISI}
    (x^{(p)} * s_{\ts})(m\ts + e) =  \left(1 - \frac{e}{\ts}\right) x_{\mN, m}^{(p)} +  \frac{e}{\ts} x_{\mN, m+1}^{(p)}.
\end{equation}
Let now $\rndq_{\ell} = \floor{\rndd_{\ell} / \ts}$ and $E_{\ell} = \rndd_{\ell} - \rndq_{\ell} \ts$ so that $\rndd_{\ell} =~ \rndq_{\ell} \ts + \rnde_{\ell}$ with $\rndq_{\ell} \in \mathbb{Z}$ and $\rnde_{\ell} \in [0, \ts)$. It follows from \eqref{eq:pilot:ISI} that we can represent $\rvecy_{\ell}^{(p)}$ as 
\begin{IEEEeqnarray}{rCl}
    \rvecy^{(p)}_\ell &=& \rndh_\ell \begin{bmatrix}\vecx^{(p)}_{\mN}(\rndq_{\ell}) & \vecx^{(p)}_{\mN}(\rndq_{\ell} + 1)\end{bmatrix} \begin{bmatrix}1 - \frac{E_{\ell}}{\ts} \\ \frac{E_{\ell}}{\ts}\end{bmatrix} \nonumber \\
     &&+ \rvecz_\ell + \rvecc_\ell, \label{eq:YpMMform}
\end{IEEEeqnarray}
where 
\begin{equation} \label{eq:xn_def_scaled}
    \vecx^{(p)}_{\mN}(q_{\ell}) = \frac{1}{\sqrt{\mN}}\ltrsqr{
    \boldsymbol{0}^{T}_{q_{\ell}}, x^{(p)}_{\mN,1}, \ldots, x^{(p)}_{\mN,\mN n_p}, \boldsymbol{0}^{T}_{\mM-q_{\ell}-\mN n_p}}^T,
\end{equation}
$\rvecz_{\ell} \sim \jpg(\boldsymbol{0}_{\mM}, \boldsymbol{\mathsf{I}}_{\mM})$ is the Gaussian vector containing the sampled noise process, and $\rvecc_{\ell}$ captures the potential interference from the data signal \eqref{eqn:cts-data}.

For simplicity, our algorithms are derived by considering the ML estimators of the parameters in question in the case where $\rvecc_{\ell} = \boldsymbol{0}_{\mM}$, i.e., where no interference from the data symbols is present. As we shall verify in Section \ref{sec:NumericalResults}, our assumption incurs minimal loss.

\subsubsection{Per-Block Synchronization for Independent Delays}\label{subsubsec:ind}
\begin{figure}
\centering
   \begin{minipage}{\columnwidth}
  \centering
  \subfloat[Per-block synchronization]{\includegraphics[width=0.7\columnwidth,keepaspectratio]{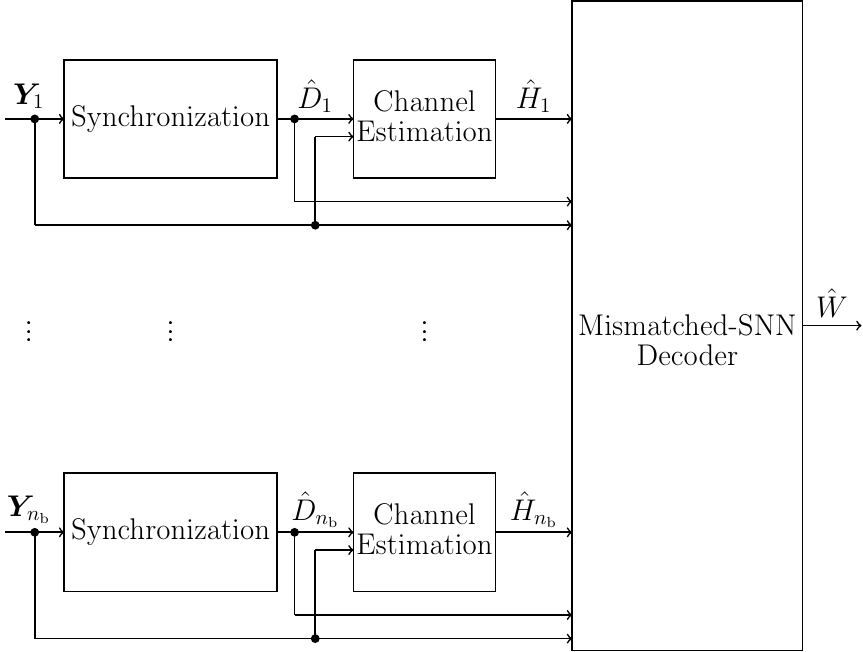} \label{fig:blockDiagram_PerBlock}}
  \end{minipage}%
  
  \begin{minipage}{\columnwidth}
  \centering
  \subfloat[Joint Synchronization]{\includegraphics[width=0.7\columnwidth,keepaspectratio]{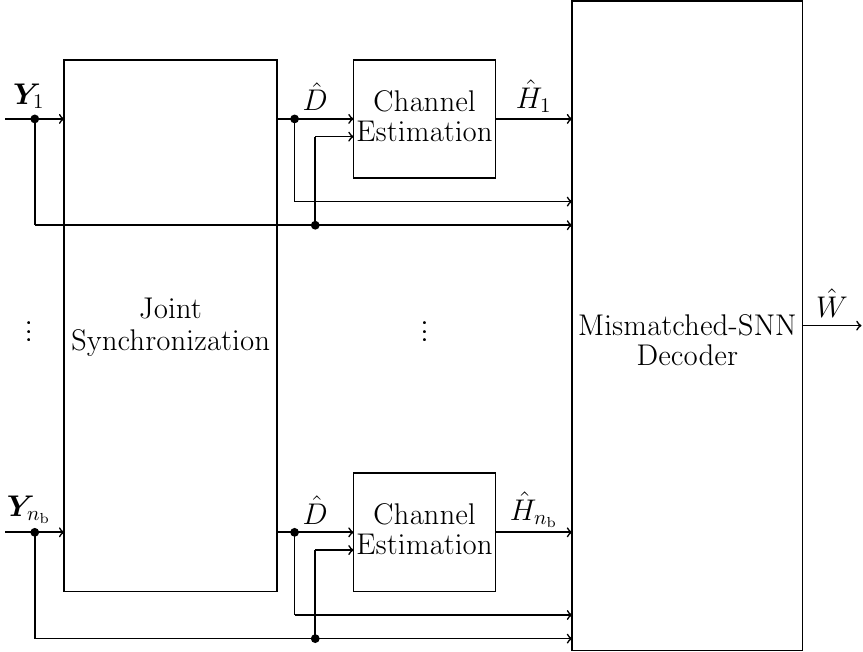} \label{fig:blockDiagram_JointSync}}
  \end{minipage}
  \caption{Block diagrams of the receiver.}
  \label{fig:BlockDiagramReceiver}
\end{figure}
In this case, as illustrated in Fig.~\ref{fig:blockDiagram_PerBlock}, the receiver uses the observation of $\rvecy_\ell^{(p)}$ for each fading block in order to obtain the estimates $\hat{\rndh}_\ell$, $\hat{\rndq}_\ell$, and $\hat{E}_\ell$. 
To do so, we pose the following minimization problem:
\begin{equation}
    \label{eq:ObjFuncSync-ind}
    [\hat{\rndh}_\ell, \hat{\rndq}_\ell, \hat{E}_\ell] = \argmin_{\bar{h},\bar{q},\bar{e}} \vecnorm{\rvecy^{(p)}_\ell - \bar{h} \vecv(\bar{q},\bar{e}) }^2,
\end{equation}
where 
\begin{equation}
    \vecv(\bar{q},\bar{e}) =  \begin{bmatrix}\vecx^{(p)}_{\mN}(\bar{q}) & \vecx^{(p)}_{\mN}(\bar{q} + 1)\end{bmatrix} \begin{bmatrix}1 - \frac{\bar{e}}{\ts} \\ \frac{\bar{e}}{\ts}\end{bmatrix}. 
\end{equation}
For any given $\rvecy^{(p)}_{\ell} = \vecy^{(p)}_{\ell}$, the ML channel estimate for each fading block $\bar{h}_{\ell}$ can be written as a function of $\bar q$ and $\bar e$ as
\begin{align}
   \hat{h}_{\ell}(\bar q, \bar e) 
   &= \argmin_{\bar{h}_{\ell}}\sum_{k=1}^{\nb} \vecnorm{\vecy^{(p)}_k - \vecv(\bar{q},\bar{e})\bar{h}_k }^2\\
   &= \argmin_{\bar{h}_{\ell}} \vecnorm{\vecy^{(p)}_\ell -  \vecv(\bar{q},\bar{e})\bar{h}_\ell}^2\\ 
   &= \frac{\vecv(\bar{q},\bar{e})^H \vecy^{(p)}_{\ell}}{||\vecv(\bar{q},\bar{e})||^2}.\label{eq:channelMLest_opt}
\end{align}
Substituting \eqref{eq:channelMLest_opt} into \eqref{eq:ObjFuncSync-ind}, we can estimate the delay parameters as
\begin{align}
    \ltrsqr{\hat{q}_{\ell},\hat{e}_{\ell}} 
    &= \argmin_{\bar{q},\bar{e}} 
    \vecnorm{\vecy_{\ell}^{(p)} - \vecv(\bar{q},\bar{e})\hat{h}_{\ell}(\bar q, \bar e) }^2\\
    &= \argmin_{\bar{q},\bar{e}}
    \vecnorm{\vecy_{\ell}^{(p)}}^2 
    - 2 \re\ltrcurley{(\vecy_{\ell}^{(p)})^H\vecv(\bar{q},\bar{e})\hat{h}_{\ell}(\bar q, \bar e)} \nonumber
    \\ &\qquad + \vecnorm{\vecv(\bar{q},\bar{e})}^2\abs{\hat{h}_{\ell}(\bar q, \bar e)}^2\\
    &= \argmin_{\bar{q},\bar{e}}
    - 2\frac{ \abs{\vecv(\bar{q},\bar{e})^{H} \vecy_{\ell}^{(p)}}^2}{\vecnorm{\vecv(\bar{q},\bar{e})}^2} 
    + \frac{\abs{\vecv(\bar{q},\bar{e})^{H} \vecy_{\ell}^{(p)}}^2}{\vecnorm{\vecv(\bar{q},\bar{e})}^2} \\
    &= \argmax_{\bar{q},\bar{e}} \frac{\abs{\vecv(\bar{q},\bar{e})^{H} \vecy_{\ell}^{(p)}}^2}{\vecnorm{\vecv(\bar{q},\bar{e})}^2}\\
    &= \argmax_{\bar{q},\bar{e}} \frac{n(\bar{q},\bar{e})}{d(\bar{q},\bar{e})},
    \label{eq:delayMLest_Per}
\end{align}
where we set
\begin{align}
    n(\bar{q},\bar{e}) &= \abs{\vecv(\bar{q},\bar{e})^{H} \vecy_{\ell}^{(p)}}^2,\\
    d(\bar{q},\bar{e}) &= \vecnorm{\vecv(\bar{q},\bar{e})}^2.       
\end{align}

For a fixed value of $\bar{q}$, both $n(\bar{q},\bar{e})$ and $d(\bar{q},\bar{e})$ are second-degree polynomials in $\bar{e}$, which implies that the objective function in~\eqref{eq:delayMLest_Per} is a rational function in $\bar{e}$. For a fixed value of $\bar{q}$, we find an extreme point of this rational function by differentiating with respect to $\bar{e}$ and setting the derivative to zero as
\begin{equation}
    \label{eq:derivEps-ind}
    \frac{\partial}{\partial \bar{e}} \frac{n(\bar{q},\bar{e})}{d(\bar{q},\bar{e})} = \frac{n'(\bar{q},\bar{e}) d(\bar{q},\bar{e}) - n(\bar{q},\bar{e}) d'(\bar{q},\bar{e}) }{d(\bar{q},\bar{e})^2} = 0. 
\end{equation}
If \eqref{eq:derivEps-ind} has a solution $\bar{e}^\star$ in the range $\left(0, t_s\right)$, then $\left(\bar{q},\bar{e}^\star\right)$ is a candidate for $\left(\hat{q}_{\ell},\hat{e}_{\ell}\right)$. We also consider the boundary points $\left(\bar{q},0\right)$ and $\left(\bar{q},t_s\right)$ as candidates, since they might be the solution of \eqref{eq:derivEps-ind} in the cases where no extreme point can be found in $\left(0, t_s\right)$, or when a minimum point (instead of a maximum) is found.

We are now ready to introduce our delay and channel estimation algorithm. Let $\mathcal{D}_{\ell}$ be the set of candidates for $\hat{d}_{\ell} = \hat{q}_{\ell}\ts + \hat{e}_{\ell}$. We construct $\mathcal{D}_{\ell}$ as follows: for each $\bar{q} \in \{0,\ldots, d_{\text{max}} / \ts \}$ we add $\bar{q}\ts$ to $\mathcal{D}_{\ell}$. Next, we find the solutions $\bar{e}^\star$ for \eqref{eq:derivEps-ind} for each $\bar{q}$. If $\bar{e}^\star$ is in the range $[0,\ts)$, we add $\bar{q}\ts + \bar{e}^\star$ to $\mathcal{D}_{\ell}$. We then find $\hat{d}_{\ell}$ as the entry in $\mathcal{D}_{\ell}$ that maximizes the objective function in~\eqref{eq:delayMLest_Per}. This estimate is then used to evaluate the channel estimate $\hat{h}_{\ell}$ from \eqref{eq:channelMLest_opt}. A pseudocode for the algorithm is given in Algorithm \ref{alg:Synchronization}.
\begin{algorithm}[t]
    \caption{Synchronization and Channel Estimation}\label{alg:synchronization:channel:estimation}
    \begin{algorithmic}
        \Require $\ts$, $d_{\max}$, $\vecy_1^{(p)}, \vecy_2^{(p)}, \ldots, \vecy_{\nb}^{(p)}$
        \Ensure A solution $\hat q$, $\hat e$, $\hat h_1, \hat h_2, \ldots, \hat h_{\nb}$ to \eqref{eq:ObjFuncSync-ind} 
        \For{$\ell = 1, 2, \ldots, \nb$}
        \State $\mathcal{D}_{\ell} \gets \{(q, 0): q = 0, 1, \ldots, \lceil d_{\text{max}} / \ts \rceil\}$
    \State \qquad \quad $\cup \{(q, \ts): q = 0, 1, \ldots, \lceil d_{\text{max}} / \ts \rceil\}$
        \For{$\bar q\in\{0,1, \ldots, \lceil d_{\text{max}} / \ts\rceil\}$} 
            \For{each root $\bar{e}^*$ to \eqref{eq:derivEps-ind} such that $\bar{e}^*\in(0,t_s)$}
                \State add $(\bar q, \bar{e}^*)$ to $\mathcal{D}_{\ell}$
            \EndFor
        \EndFor
        \State $(\hat{q}_{\ell}, \hat{e}_{\ell})\gets$ element in $\mathcal{D}_{\ell}$ that maximizes $\eqref{eq:delayMLest_Per}$ 
            \State $\hat h_\ell \gets \vecv(\hat{q},\hat{e})^H \vecy^{(p)}_{\ell} / ||\vecv(\hat{q},\hat{e})||^2$
        \EndFor    
    \end{algorithmic}
    \label{alg:Synchronization}
\end{algorithm}
\subsubsection{Joint Synchronization for Fully Dependent Delays}\label{subsubsec:joint}
In this case, since the delay is same for all fading blocks (i.e., $\rndd_{\ell} = \rndd$ for all $\ell$), the receiver can use the observations of $\{\rvecy_\ell^{(p)}\}_{\ell=1}^{\nb}$ jointly, as illustrated in Fig. \ref{fig:blockDiagram_JointSync}. These observations are used to estimate the channel gains $\hat{\rvech} = [\hat{\rndh}_1,\ldots,\hat{\rndh}_{\nb}]^{T}$ and delay as 
\begin{equation}
    \label{eq:ObjFuncSync}
    [\hat{\rvech}, \hat{\rndq}, \hat{E}] = \argmin_{\bar{\vech},\bar{q},\bar{e}} \sum_{\ell=1}^{\nb} \vecnorm{\rvecy^{(p)}_\ell - \bar{h}_\ell \vecv(\bar{q},\bar{e})}^2.
\end{equation}
Following the same steps leading to \eqref{eq:delayMLest_Per}, we obtain
\begin{align}
    \ltrsqr{\hat{q},\hat{e}} &= \argmax_{\bar{q},\bar{e}} \sum_{\ell=1}^{\nb} \frac{\abs{\vecv(\bar{q},\bar{e})^{H} \vecy_{\ell}}^2}{\vecnorm{\vecv(\bar{q},\bar{e})}^2}.
                             \label{eq:delayMLest}
\end{align}

The numerator and denominator of \eqref{eq:delayMLest} have the same polynomial structure as \eqref{eq:delayMLest_Per}, and the same synchronization and channel estimation algorithm described in Section~\ref{subsubsec:ind} can be used, but now with only one set of candidates $\mathcal{D}$. 

\subsection{Codeword Decoding Phase} \label{subsec:code}
The codeword decoding phase is based on a mismatch-decoding approach, where the delay and channel estimates returned by the algorithms described in Section \ref{subsec:synch} are treated as perfect. 
The input-output relationship for the $k$th symbol in the $\ell$th block is\footnote{We will omit the superscript $^{(d)}$ in the remainder of the paper to keep the notation compact.}
\begin{equation}
    \rndy_{k,\ell} = \ltrp{\rndy_{\ell} * s_{\tp}}(k t_p + n_p t_p +\hat{\rndd}_\ell), \quad k=1, 2, \ldots, \ns.
    \label{eq:Y_kl_1}
\end{equation}

In the decoding process, the receiver seeks the codeword in the codebook $\mathcal{C}$ closest to the received signal after scaling each subcodeword with the corresponding channel estimates. Hence, given the received vector and the channel estimates, the decoded codeword $\hat{\vecx}=[\hat{\vecx}_{1}^{T},\dots, \hat{\vecx}_{\nb}^{T}]^T$ is determined as
\begin{equation}
    \hat{\vecx} = \argmin_{ 
        \substack{
        \bar{\vecx} = [\bar{\vecx}_{1}^{T},\ldots,\bar{\vecx}_{\nb}^{T}]^{T} \\ 
        \bar{\vecx}  \in\setC} } \sum_{\ell=1}^{\nb} \vecnorm{\vecy_{\ell}
			      -\hat{h}_{\ell}\bar{\vecx}_\ell }^{2}, \label{eq:snn_dec}
\end{equation}
where $\vecy_{\ell} = [y_{1,\ell}, \ldots, y_{\ns,\ell}]^T$.
This decoder, known as the \textit{mismatched SNN decoder} \cite{Scarlett2014}, coincides with the ML decoder only when the receiver has perfect channel state information, i.e., $\hat{h}_{\ell} = h_{\ell}$
for $\ell = 1,\ldots,\nb$, and perfect synchronization. This decoder, although not optimal, is practically relevant and the analysis of its finite-blocklength error probability is tractable \cite{Ostman2019}.

\section{A Non-Asymptotic Upper Bound on The Error Probability}
\label{sec:RCUsSection}
Let $\hat{\rndw}$ be the estimate, produced by the decoder, of the transmitted message $\rndw$. We define the packet error probability as $\epsilon\sub{pep} = \Prob[\hat{\rndw} \neq \rndw]$. We can express $\epsilon\sub{pep}$ as
\begin{equation}\label{eqn:error-bound}
    \epsilon_{\text{pep}} = \Prob\ltrsqr{\abs{\hat{\rndd}_{\ell} - \rndd_{\ell}} \leq \tp} \epsilon_1 +  \Prob\ltrsqr{\abs{\hat{\rndd}_{\ell} - \rndd_{\ell}} > \tp} \epsilon_2,
\end{equation}
where $\epsilon_1$ and $\epsilon_2$ are the probability of erroneous packet decoding when the synchronization is off by less than and more than one symbol, respectively. When evaluating $\epsilon_{\text{pep}}$, we will assume, for simplicity, that the decoder cannot decode the packet when synchronization is off more than one symbol, i.e., we will upper-bound $\epsilon\sub{pep}$ in \eqref{eqn:error-bound} by setting $\epsilon_2 = 1$.

In the next section, we will present an RCUs bound for $\epsilon_1$ and its corresponding saddlepoint approximation.
\subsection{The RCUs Finite-Blocklength Bound}
Like most of the achievability results in information theory, the RCUs bound is obtained by a random-coding argument. This means that instead of analyzing the performance of a particular code, we evaluate the average error probability averaged over a randomly constructed ensemble of codebooks. 
In this paper, we consider an i.i.d. discrete ensemble in which each symbol of every codeword is drawn independently (and uniformly) from a constellation set $\setU$ with $u$ elements (e.g., $u=2$ for BPSK) and power $\rho$. Although potentially suboptimal, this choice is practically relevant and allows us to evaluate the RCUs bound efficiently via a saddlepoint approximation.

When the synchronization is off by less than one symbol (i.e., $|\hat{\rndd}_{\ell} - \rndd_{\ell}| \leq \tp$), the input-output relation in \eqref{eq:Y_kl_1} can be stated as
\begin{equation}
    \rndy_{k,\ell} = \rndh_\ell \ltrp{\Delta_\ell X_{k,\ell} + (1-\Delta_\ell) X_{k+\Lambda_\ell,\ell}} + \rndz_{k,\ell} , \label{eq:dataSignal}
\end{equation}
where we define
\begin{align}
\Delta_\ell &= 1-\abs{\hat{\rndd}_\ell - \rndd_\ell}/ \tp\\
\Lambda_\ell &= \text{sign}(\hat{\rndd}_\ell - \rndd_\ell),
\end{align}
and $\rndz_{k,\ell}$ are \iid zero-mean, unit-variance, complex Gaussian random variables. 
Note that the lack of perfect synchronization, i.e., $\Delta_\ell \neq 1$, yields intersymbol interference. In \eqref{eq:dataSignal}, the data symbol $\rndx_{k,\ell}$ and the intersymbol interference $\rndx_{k+\Lambda_{\ell},\ell}$ are scaled by $\Delta_\ell$ and $1-\Delta_{\ell}$, respectively, both of which are independent of $\Lambda_{\ell}$. Since data symbols are also independent of each other, $\Lambda_{\ell}$ has an impact on the error probability only through whether the last symbol of a subcodeword i.e. $\rndx_{\ns,\ell}$ is affected by intersymbol interference or not.\footnote{$\rndx_{1,\ell}$ is affected by intersymbol interference when $\Lambda_{\ell} = -1$ due to pilot symbols.} Specifically, when $\Lambda_{\ell} = -1$, $\rndx_{\ns-1,\ell}$ interferes with $\rndx_{\ns,\ell}$. Conversely, when $\Lambda_{\ell} = 1$, $\rndx_{\ns,\ell}$ is free from intersymbol interference. Even if the impact of $\Lambda_\ell$ on the packet error probability is expected to be minute, we assume the worst case here and let $\Lambda_{\ell} = -1$ for the remainder of the paper. By choosing the worst-case scenario for our analysis, we ensure that the achievability bound we shall present next still holds as an achievability bound.

For this setup, the RCUs achievability bound $\epsilon_{\text{ub}}$ on $\epsilon_1$ is given by~\cite{Martinez2011}
\begin{equation}
    \label{eq:RCUsVec}
    \epsilon_1 \leq \epsilon_{\text{ub}}
\end{equation}
with
\begin{equation}
    \label{eq:rcus-average}
    \epsilon\sub{ub} = \Ex{\rvech,\hat{\rvech}, \boldsymbol{\Delta}}{\epsilon_{\text{ub}} \ltrp{\rvech,\hat{\rvech}, \boldsymbol{\Delta}}}
\end{equation}
where $\rvech = [\rndh_{1}, \ldots, \rndh_{\nb}]^T$, $\boldsymbol{\Delta} = [\Delta_{1}, \ldots, \Delta_{\nb}]^T$, and
\begin{IEEEeqnarray}{rCl}
    \nonumber
    \epsilon_{\text{ub}}\ltrp{\vech,\hatVech,\boldsymbol{\delta}}\nonumber = \Prob\biggl[&& \frac{\log \Upsilon}{\nc\nb} + \frac{1}{\nc\nb} 
  \sum_{\ell=1}^{\nb} \sum_{k=1}^{\ns}
   \infden\ltrp{\rndx_{k,\ell}; \rndy_{k,\ell}, \hat{h}_{\ell}} \\ 
    && \qquad
   \leq \rate \bigg|   \rvech=\vech, \hat{\rvech}=\hatVech, \boldsymbol{\Delta} = \boldsymbol{\delta} \biggr]. \label{eq:RCUsSym-cond}
\end{IEEEeqnarray}
Here, $\Upsilon$ is a random variable that is uniformly distributed on $[0,1]$ and independent of all other quantities,
and $\infden(x,y,\hat{h})$ is the so-called \emph{generalized information density}, which is defined as \cite{Martinez2011}
\begin{equation}
    \label{eq:InfoDens}
    \infden(x;y,\hat{h}) = \log \frac{e^{-s\abs{y - \hat{h} x}^2}}{\Ex{\bar{\rndx}}{e^{-s\abs{y - \hat{h}\bar{\rndx}}^2}}},
\end{equation}
where $\bar{\rndx}$ is independent of all other random variables and is drawn uniformly from the constellation and $s > 0$ is a parameter that can be optimized to obtain a tighter bound. Note that \eqref{eq:RCUsSym-cond} can be evaluated (at least numerically) for any realization of $\Delta$, $\rndh$ and $\hat{\rndh}$, regardless of their underlying distribution. Consequently, the RCUs bound can be evaluated for all channel and delay estimation algorithms.


No closed-form expression for the RCUs bound~\eqref{eq:RCUsSym-cond} is in general available. Numerical methods to evaluate it, such as Monte-Carlo simulations, can be time consuming due to the low target error probabilities of interest in URLLC.
Next, we introduce a saddlepoint approximation on \eqref{eq:RCUsSym-cond} that allows for an efficient computation of \eqref{eq:RCUsSym-cond}.

\section{A Saddlepoint Approximation on the Conditional Error Probability in \eqref{eq:RCUsSym-cond}}
The saddlepoint method is a well-established tool to obtain accurate approximations of tail
probabilities involving sum of random variables. 
Unfortunately, none of the saddlepoint approximations for the RCUs bounds reported previously in the
literature apply to the setup considered in this paper.
Indeed, differently from \cite{Kislal2023,ostman2020}, in our setup,
$\{\rndy_{k,\ell}\}_{k=1}^{\ns}$ are not conditionally independent given $\rvech=\vech$,
$\hat{\rvech}=\hatVech$, and $\boldsymbol{\Delta} =\boldsymbol{\delta}$, due to intersymbol interference (see \eqref{eq:dataSignal}). 
As a result, the random variables $\{\infden(\rndx_{k,\ell}; \rndy_{k,\ell}, \hat{h}_{\ell})\}_{k=1}^{\ns}$ are also not conditionally independent.
This prevents us from using the saddlepoint expansion reported, e.g., in~\cite[Thm.~1]{Kislal2023}.

In the case of independent delays, the $\{\rndy_{k,\ell}\}_{\ell=1}^{\nb}$ are independent for
each $\ell$, and
since delay and channel estimation are performed per block, this implies that the random variables
$\left\{\sum_{k=1}^{\ns}
\infden\ltrp{\rndx_{k,\ell}; \rndy_{k,\ell}, \hat{\rndh}_{\ell}}\right\}_{\ell=1}^{\nb}$ are also
independent.
This would in principle enable us to derive a saddlepoint expansion directly on the $\epsilon\sub{ub}$
in~\eqref{eq:rcus-average}, similar to the one reported
in~\cite[Thm.~2]{Kislal2023}.\footnote{We would, however, still need to deal with the intersymbol-interference within each block.}
However, this approach cannot be applied to the case of fully dependent delays.

To obtain a saddlepoint approximation that applies to both the cases of fully dependent delays
and independent delays, we exploit the property that, in both cases, the random variables
$\left\{\sum_{k=1}^{\ns}\imath_s(X_{k,\ell};Y_{k,\ell},
\hat{h}_\ell)\right\}_{\ell=1}^{n\sub{b}}$ are conditionally independent, given $\rvech=\vech$,
$\hat{\rvech}=\hat{\vech}$, and $\boldsymbol{\Delta}=\boldsymbol{\delta}$. 
This allows us to perform a saddlepoint expansion of the conditional error probability
$\epsilon_{\text{ub}}\ltrp{\vech,\hatVech,\boldsymbol{\delta}}$ in~\eqref{eq:RCUsSym-cond} with respect
to the number of blocks $n\sub{b}$.
Specifically, let us fix $\boldsymbol{\delta}$, $\vech$, and $\hat{\vech}$, and denote for
convenience 
\begin{IEEEeqnarray}{rCl}\label{eq:def_Il}
  I_{\ell} = \sum_{k=1}^{\ns} \infden(X_{k,\ell};Y_{k,\ell},
\hat{h}_\ell),
\end{IEEEeqnarray}
where we let, with a slight abuse of notation, for the remainder of this section
\begin{align}
    &\infden(\rndx_{k,\ell}; \rndy_{k,\ell}, \hat{h}_{\ell}) \nonumber \\ & =
    \log \frac{e^{-s
    \abs{h_\ell \ltrp{\delta_\ell X_{k,\ell} + (1-\delta_\ell) X_{k-1,\ell}} + \rndz_{k,\ell} - \hat{h}_\ell X_{k,\ell}}^2}}{\Ex{\bar{\rndx}}{e^{-s\abs{h_\ell \ltrp{\delta_\ell X_{k,\ell} + (1-\delta_\ell) X_{k-1,\ell}} + \rndz_{k,\ell} - \hat{h}_\ell\bar{\rndx}}^2}}},
    \label{eq:is:conditional}
\end{align}
which is a function of the random variables $X_{k,\ell}$, $X_{k-1,\ell}$, and $\rndz_{k,\ell}$ (and is parameterized by $h_\ell$, $\hat h_\ell$, and $\delta$). 

As we shall see, to obtain the desired saddlepoint expansion, we need to solve the following two issues.
\begin{itemize}
	\item In our setup, the $\{I_{\ell}\}_{\ell=1}^{\nb}$ are conditionally independent, but, differently from the variables appearing in previous works, e.g.,~\cite{ostman2020,Kislal2023}, they are not identically distributed. 
	\item To apply the saddlepoint expansion, we need to be able to compute efficiently (ideally, in closed-form) the moment-generating function (MGF) of each $I_{\ell}$, which requires us to tackle the dependence between the random variables in the sum on the right-hand side of~\eqref{eq:def_Il}, caused by intersymbol interference.
\end{itemize}

\subsection{Saddlepoint Expansion for Sum of Independent but not Identically Distributed Random
Variables}
We next present a general result that will allow us to obtain a saddlepoint approximation on 
$\epsilon_{\text{ub}}\ltrp{\vech,\hatVech,\boldsymbol{\delta}}$ in~\eqref{eq:RCUsSym-cond}.
Let $\{R_\ell\}_{\ell=1}^{n}$ be a family of independent, but not necessarily identically
distributed, nonlattice random variables.\footnote{A random variable $R_{\ell}$ is said to be lattice if it is supported on the points $b, b\pm h, b\pm 2h,\dots$ for some $b \in \mathbb{R}$ and $h \in \mathbb{R}$. A random variable that is not
lattice is said to be nonlattice.}
Let us denote by $\varphi_{\ell}(\zeta)=\Ex{}{e^{-\zeta R_{\ell}}}$ the MGF of
$-R_{\ell}$ and by $\kappa_{\ell}(\zeta)=\log \varphi_{\ell}(\zeta)$ its cumulant generating function (CGF).
Let us also set $\kappa(\zeta)=\sum_{\ell=1}^{n}\kappa_{\ell}(\zeta)$ as well as 
\begin{IEEEeqnarray}{rCl}
    \mu(\zeta)&=& \frac{1}{n} \odv{\kappa(\zeta)}{\zeta}    \\ 
    \sigma^{2}(\zeta)&=& \frac{1}{n} \odv[order=2]{\kappa(\zeta)}{\zeta} .
\end{IEEEeqnarray}

In the next theorem, we present a saddlepoint approximation for the probability that $\log \Upsilon +
\sum_{\ell=1}^{n} R_{\ell}$ does not exceed $n\rate$.  

\begin{thm}\label{thm:saddlepoint-generic}
  Suppose that there exists a $\zeta_{0}>0$ such that 
  \begin{equation}\label{eq:fourth-moment-constraint}
      \sup_{\abs{\zeta}<\zeta_{0}} \abs{\odv[order=4]{\varphi_{\ell}(\zeta)}{\zeta}}<\infty, \quad \forall
      \ell\in\{1,\dots,n\}   
  \end{equation}
  and also positive constants $m\sub{l}\leq m\sub{u}$ such that 
 \begin{equation}\label{eq:second moment constraint}
     m\sub{l} \leq \sigma^{2}(\zeta) \leq m\sub{u}
 \end{equation}
 holds for all $n \in \mathbb{N}$ and for all $\abs{\zeta}\leq\zeta_{0}$.
 Assume that there exists a $\zeta \in [-\zeta_{0},\zeta_{0}]$ satisfying $-\mu(\zeta)=\rate$.
 If $\zeta \in [0,1]$ then 
 \begin{align}
     &\Prob\ltrsqr{\log \Upsilon + \sum_{\ell=1}^{n}R_{\ell} \leq n\rate} \nonumber \\ 
     & = e^{\kappa(\zeta) -  n\zeta \mu(\zeta) } \ltrsqr{ e^{\frac{\beta_\zeta^2}{2}} Q(\beta_\zeta) + e^{\frac{\beta_{1-\zeta}^2}{2}}
     Q(\beta_{1-\zeta}) + \landauo\ltrp{\frac{1}{\sqrt{n}}} }  \label{eq:SPMethod1}
 \end{align}
where $\beta_a = a \sqrt{n\sigma^2(\zeta) }$.
If $\zeta>1$, then 
\begin{align}
   &\Prob\ltrsqr{\log \Upsilon + \sum_{\ell=1}^{n}R_{\ell} \leq n\rate } \nonumber \\ 
   &=  e^{\kappa(1) -
n\mu(\zeta)}  \ltrsqr{\Psi_{n}(1,1)-\Psi_{n}(0,-1) + \landauO\ltrp{\frac{1}{\sqrt{n}}}}\label{eq:SPMethod2}
\end{align}
where 
\begin{align}
    \Psi_{n}(a,b) &= e^{na[-\mu(1)-\rate + \sigma^{2}(1)/2]} \nonumber \\
    & \quad \times Q\ltrp{a \sqrt{n\sigma^{2}(1)}-b
\frac{n(\mu(1)+\rate)}{\sqrt{n\sigma^{2}(1)}}}. \label{eq:SP_Psi2}
\end{align}
Finally, if $\zeta<0$, then
\begin{align}
   &\Prob\ltrsqr{\log \Upsilon + \sum_{\ell=1}^{n}R_{\ell} \leq n\rate } = 1- \ltrp{e^{\kappa(\zeta) -  n\zeta \mu(\zeta) }}  \nonumber \\ 
   & \qquad \times
   \ltrsqr{e^{\frac{\beta_{-\zeta}^2}{2}} Q(\beta_{-\zeta}) - e^{\frac{\beta_{1-\zeta}^2}{2}}
   Q(\beta_{1-\zeta}) + \landauO\ltrp{\frac{1}{\sqrt{n}}} }.\label{eq:SPMethod3}
\end{align}
\end{thm}
\begin{IEEEproof}
    To prove Theorem~\ref{thm:saddlepoint-generic} one has to combine the steps in~\cite[App.
    I]{Lancho2020} with the ones in~\cite[App. E]{Scarlett2014}.
    Furthermore, one has to replace~\cite[Lem.~8]{Lancho2020} with~\cite[Thm.~1,
    Sec.~XVI.6]{Feller1971}.
\end{IEEEproof}

To use Theorem~\ref{thm:saddlepoint-generic} in order to obtain a saddlepoint expansion
of~\eqref{eq:RCUsSym-cond}, one has to replace the $\{R_{\ell}\}$ with the $\{I_{\ell}\}$ defined
in~\eqref{eq:def_Il}, $n$ with $\nb$, and $\rate$ by $\nc\rate$. 
Verifying that conditions~\eqref{eq:fourth-moment-constraint} and~\eqref{eq:second moment constraint}
hold in our scenario is challenging. 
We shall return to this point in Section \ref{sec:NumericalResults}. 
The saddlepoint approximation is then obtained by neglecting the $\landauo(\cdot)$ and
$\landauO(\cdot)$ terms in~\eqref{eq:SPMethod1},~\eqref{eq:SPMethod2}, and~\eqref{eq:SPMethod3}.
A pseudocode for computing $\epsilon\sub{ub}$ in \eqref{eq:RCUsVec} using this approximation is provided in Algorithm~\ref{alg:cap}.
\begin{algorithm}[t]
\caption{Saddlepoint Approximation}\label{alg:cap}
\begin{algorithmic}[1]
\State \textbf{Input} : $\rho, \rate,\ns,\np,\nb,\mN\sub{MC}$
\State \textbf{Output} : $\epsilon\sub{ub}$
\State Draw $(\vech_i, \hat{\vech}_i, \boldsymbol{\delta}_i)$ for $i = 1, \ldots, \mN\sub{MC}$
\For{every $(\vech_i, \hat{\vech}_i, \boldsymbol{\delta}_i)$}
    \State Solve $\rate = \frac{-\mu(\zeta)}{\nc}$ for $\zeta$, denote solution by $\zeta^*$
    \State Compute $\kappa(\zeta^*)$, $\kappa''(\zeta^*)$
    \State Compute $\epsilon\sub{ub}(\vech_i,\hat{\vech}_i, \boldsymbol{\delta}_i)$ using Theorem \ref{thm:saddlepoint-generic}
\EndFor
\State $\epsilon\sub{ub} \gets \frac{1}{\mN\sub{MC}} \sum_{i=1}^{\mN\sub{MC}} \epsilon\sub{ub}(\vech_i,\hat{\vech}_i, \boldsymbol{\delta}_i) $
\end{algorithmic}
\label{alg:seq}
\end{algorithm}

\subsection{Closed-Form Evaluation of the Moment Generating Function}
It follows from the pseudocode provided in Algorithm~\ref{alg:cap} that a key step in the computation of
the saddlepoint approximation is the evaluation of the moment-generating function
$\varphi_{\ell}(\zeta)=~\Ex{}{e^{-\zeta I_{\ell}}}$,
for $\ell=1,\dots,\nb$, which is required to determine $\kappa(\zeta)$ and its first and second derivatives.
As shown in~\eqref{eq:def_Il}, each $I_{\ell}$ consists of the sum of $\ns$ random
variables. 
However, these random variables are not independent because of the intersymbol interference
caused by errors in the estimation of the propagation delay (see~\eqref{eq:is:conditional}).
Inspired by \cite[Ch. 9]{jensen95-a}, we shall next exploit the Markovian structure of the
dependence between these random variables to provide an effective method to evaluate
$\varphi_{\ell}(\zeta)$.

Let $\setU$ denote the set of constellation points of cardinality $u$.
It will turn out convenient to associate to every pair\footnote{When considering pulse shapes that create more intersymbol interference, $B_{k,\ell}$ may need to be associated with a larger number of adjacent symbols, leading to a more involved state-space-evolution analysis.} $(X_{k-1,\ell},X_{k,\ell})\in \setU^{2}$ of consecutive transmitted symbols for $k=1,\dots,\ns$, with an index $B_{k,\ell}\in\{1,\dots,u^{2}\}$.
In Table~\ref{tab:table1}, we present one such association for the case of BPSK constellation.
\begin{table}
    \caption{A possible mapping between $(X_{k-1,\ell},X_{k,\ell})$ and $B_{k,\ell}$ for BPSK
    constellation; here, $x^{(1)}=-\sqrt{\rho}$ and $x^{(2)}=~\sqrt{\rho}$.}
    \label{tab:table1}
    \centering
    \begin{tabular}{@{}ccc@{}}
       $B_{k,\ell}$  & $X_{k-1,\ell}$ & $X_{k,\ell}$  \\
       \hline
         1 & $x^{(1)}$ & $x^{(1)}$   \\ 
         2 & $x^{(1)}$ & $x^{(2)}$   \\ 
         3 & $x^{(2)}$ & $x^{(1)}$   \\ 
         4 & $x^{(2)}$ & $x^{(2)}$  
    \end{tabular}
\end{table}
Note that the index $B_{1,\ell}$ depends on the pair  $(X_{0,\ell},X_{1,\ell})$, where $X_{0,\ell}$
stands for the last pilot symbol transmitted before the subcodeword. 
We shall assume for simplicity that, as for the data symbols, this pilot symbol is drawn uniformly and independently from the constellation~$\setU$.

We note that for all possible values of $b_{k,\ell}, b_{k-1,\ell}, \ldots, b_{1,\ell}$, we have that
\begin{align}
\label{eq:dk1}
 &\Prob\ltrsqr{\rndb\spk = b_{k,\ell}\given \rndb_{k-1,\ell} = b_{k-1,\ell}, \ldots, \rndb_{1,\ell} = b_{1,\ell}} 
 \nonumber \\ 
 &= \Prob[\rndb\spk = b_{k,\ell} \given \rndx_{k-1,\ell} =x_{k-1,\ell}, \rndx_{k-2,\ell}=x_{k-2,\ell}] \\
 & = \Prob\ltrsqr{\rndb\spk = b_{k,\ell} \given \rndb\spkk = b_{k-1,\ell}},
\end{align}
which implies that the sequence $\{B_{k,\ell}\}$ is a Markov chain in~$k$.

Let $\{(x_{i1},x_{i2})\}_{i=1}^{u^{2}}$ denote the set of all possible pairs of consecutively
transmitted constellation symbols. 
In the remainder of the paper, we will use the convention that $B_{k,\ell}=i$ means that
$X_{k-1,\ell}=x_{i1}$ and $X_{k,\ell}=x_{i2}$.
Using this notation, we can express each entry $[\matP]_{ij}$ of the $u^{2}\times u^{2}$ transition matrix\footnote{
The transition matrix does not depend on the index $\ell$; hence, this index is omitted.} $\matP$ as 
\begin{align}
    [\matP]_{ij} &= \Prob\ltrsqr{\rndb_{k+1,\ell} = j \ggiven \rndb_{k,\ell} = i}\\
    &= \Prob[ \rndx_{k+1,\ell} = x_{j2}, \rndx_{k,\ell} = x_{j1} \nonumber \\ 
    & \qquad \qquad \given \rndx_{k,\ell} = x_{i2}, \rndx_{k-1,\ell} = x_{i1} ] \\ 
     &= \Prob\ltrsqr{\rndx_{k+1,\ell} = x_{j2}} \ind{x_{j1} = x_{i2}}\\
     &= \frac{1}{u}\ind{x_{j1} = x_{i2}}.
\end{align} 
It also follows by our assumptions that the initial state $B_{1,\ell}$ is chosen uniformly at random
from the set $\{1,\dots, u^{2}\}$.

Let the conditional moment generating function of $-\infden(\rndx_{k,\ell};\rndy_{k,\ell}, \hat{h}_{\ell})$, given $B_{k,\ell}=j_k$, be denoted as
%
\begin{equation}\label{eq:varphi0j}
    \varphi_{j_k,\ell}(\zeta) = \Ex{}{e^{-\zeta \infden\ltrp{\rndx_{k,\ell}; \rndy_{k,\ell},
    \hat{h}_\ell}} \ggiven \rndb_{k,\ell}=j_k}.  
\end{equation}
Note that the expectation in~\eqref{eq:varphi0j} is only with respect to the additive noise
$\rndz_{k,\ell}$ in~\eqref{eq:dataSignal}. 
Thus, for all $j_1, \ldots, j_{\ns} \in~ \{1, \ldots, u^2\}$ we have
\begin{align}
    \varphi_{j_k,\ell}(\zeta) &= \Exop\Big[e^{-\zeta \infden\ltrp{\rndx_{k,\ell}; \rndy_{k,\ell},
    \hat{h}_\ell}} \nonumber\\ 
    & \qquad \qquad \qquad \Big| \rndb_{1,\ell}=j_1, \ldots, \rndb_{\ns,\ell}=j_{\ns}\Big],
\end{align}
and the conditional MGF of $-I_{\ell}$ can therefore be stated as
\begin{align}
    &\Ex{}{e^{-\zeta \sum_{k=1}^{\ns}\infden\ltrp{\rndx_{k,\ell}; \rndy_{k,\ell},
    \hat{h}_\ell}} \ggiven \rndb_{1,\ell}=j_1, \ldots, \rndb_{\ns,\ell}=j_{\ns}}\notag\\
    &= \prod_{k=1}^{\ns} \Ex{}{e^{-\zeta\infden\ltrp{\rndx_{k,\ell}; \rndy_{k,\ell},
    \hat{h}_\ell}} \ggiven \rndb_{1,\ell}=j_1, \ldots, \rndb_{\ns,\ell}=j_{\ns}} \label{eq:Cond_MGF_step2}\\
    &= \prod_{k=1}^{\ns} \varphi_{j_k,\ell}(\zeta)
\end{align}
where \eqref{eq:Cond_MGF_step2} follows since $\{\rndz_{k,\ell}\}_{k=1}^{\ns}$ are \iid.
We can remove the conditioning using the law of total expectation as
\begin{IEEEeqnarray}{rCl}
    \varphi_{\ell}(\zeta) &=& \sum_{j_1=1}^{u^2}\cdots \sum_{j_{\ns}=1}^{u^2} 
    \left(\prod_{k=1}^{\ns} \varphi_{j_k,\ell}(\zeta)\right) \nonumber \\ && \times \Prob[\rndb_{1,\ell}=j_1, \ldots, \rndb_{\ns,\ell}=j_{\ns}]. \label{eq:MGF}
\end{IEEEeqnarray}
Evaluating this expression requires us to sum $u^{2\ns}$ terms, which is computationally infeasible except for very small $\ns$. We next exploit the Markov structure to find an efficient method to compute $\varphi_{\ell}(\zeta)$.

To this end, let $\matP_{\ell}(\zeta)$ be the $u^{2}\times u^{2}$ matrix with elements $[\matP_{\ell}(\zeta)]_{mn} = [\matP]_{mn}\varphi_{n,\ell}(\zeta)$ for $m,n \in\{1, \ldots, u^2\}$.
The entry in position\footnote{The indices $j_1$ and $j_{\ns}$ can be chosen arbitrarily, independent of their subscripts. We select these subscripts to keep the following derivations compact.} $(j_1, j_{\ns})$ of the matrix $\matP_{\ell}(\zeta)^{\ns-1}$ can be (for $\ns\ge~ 3$)
expressed as
\begin{align}
 &\left[\matP_{\ell}(\zeta)^{\ns-1}\right]_{j_1j_{\ns}} \nonumber \\
      &= \sum_{j_2=1}^{u^2}\cdots \sum_{j_{\ns-1}=1}^{u^2}
        \left[\matP_{\ell}(\zeta)\right]_{j_1j_2}\cdots
        \left[\matP_{\ell}(\zeta)\right]_{j_{\ns-1}j_{\ns}} \\
     &= \sum_{j_2=1}^{u^2}\cdots \sum_{j_{\ns-1}=1}^{u^2}
        \ltrp{\prod_{k=2}^{\ns} \varphi_{j_k,\ell}(\zeta)} \nonumber \\
     & \qquad \qquad \qquad \qquad \qquad \times \left[\matP\right]_{j_1j_2}\cdots \left[\matP\right]_{j_{\ns-1}j_{\ns}}. \label{eq:conditional:MGF:matrix:power}
\end{align}
Now, since
\begin{align}
    &\left[\matP\right]_{j_1j_2}\cdots
        \left[\matP\right]_{j_{\ns-1}j_{\ns}}\notag\\
        & = \prod_{k=2}^{\ns} \Prob[B_{k, \ell} = j_k\mid B_{k-1, \ell} = j_{k-1}, \ldots, B_{1, \ell} = j_{1}]\\
        & = \Prob[B_{\ns, \ell} = j_{\ns}, \ldots,  B_{2, \ell} = j_{2} \mid  B_{1, \ell} = j_{1}]\\
        & = \Prob[B_{\ns-1, \ell} = j_{\ns-1}, \ldots,  B_{2, \ell} = j_{2} \mid  B_{\ns, \ell} = j_{\ns}, \nonumber \\  
        & \qquad \qquad \qquad \quad  B_{1, \ell} = j_{1}]\Prob[B_{\ns, \ell} = j_{\ns}\mid B_{1, \ell} = j_{1}]\\
        & = \Prob[B_{\ns-1, \ell} = j_{\ns-1}, \ldots,  B_{2, \ell} = j_{2} \mid  B_{\ns, \ell} = j_{\ns},  \nonumber \\ 
        & \qquad \qquad \qquad \qquad \qquad \qquad \quad  B_{1, \ell} = j_{1}][\matP^{\ns-1}]_{j_{1}, j_{\ns}}
\end{align}
it follows (from the law of total expectation) that \eqref{eq:conditional:MGF:matrix:power} simplifies to
\begin{align}
 \left[\matP_{\ell}(\zeta)^{\ns-1}\right]_{j_1j_{\ns}} 
    &= \Exop\Big[e^{-\zeta \sum_{k=2}^{\ns} i_{s}(X_{k,\ell};Y_{k,\ell},
		\hat{h}_{\ell}) } \nonumber \\ 
  & \Big| B_{\ns,\ell}=j_{\ns},  B_{1,\ell}=j_1\Big]
		\left[\matP^{\ns-1}\right]_{j_{1}j_{\ns}}.
\end{align}
We can now form an expression for $\varphi_{\ell}(\zeta)$ as
\begin{align}
    \varphi_{\ell}(\zeta) 
    &= \Ex{}{e^{-\zeta I_{\ell}}} \\ 
	&= \Ex{}{e^{-\zeta \infden\ltrp{\rndx_{1,\ell}; \rndy_{1,\ell},
		\hat{h}_\ell}}e^{-\zeta \sum_{k=2}^{\ns} \infden\ltrp{\rndx_{k,\ell}; \rndy_{k,\ell}, \hat{h}_\ell}}} \\
	&= \sum_{j_1=1}^{u^{2}} \Exop\Big[e^{-\zeta\infden\ltrp{\rndx_{1,\ell};
			 \rndy_{1,\ell}, \hat{h}_\ell}} e^{-\zeta \sum_{k=2}^{\ns}
			 \infden\ltrp{\rndx_{k,\ell}; \rndy_{k,\ell}, \hat{h}_\ell}} \nonumber \\  
    & \qquad \qquad \qquad \qquad \qquad \Big| \rndb_{1,\ell} = j_1\Big]\Prob[B_1=j_1]\label{eq:phi:LTE}\\
	&= \sum_{j_1=1}^{u^{2}} \frac{1}{u^2} \Ex{}{e^{-\zeta\infden\ltrp{\rndx_{1,\ell};
			 \rndy_{1,\ell}, \hat{h}_\ell}} \ggiven \rndb_{1,\ell} = j_1} \nonumber \\
		&\qquad \times\Ex{}{e^{-\zeta \sum_{k=2}^{\ns}
			 \infden\ltrp{\rndx_{k,\ell}; \rndy_{k,\ell}, \hat{h}_\ell}} \ggiven
		     \rndb_{1,\ell} = j_1} \label{eq:phi:conditional:independence}\\ 
	&= \sum_{j_1=1}^{u^{2}} \frac{\varphi_{j_1, \ell}(\zeta)}{u^2}  \nonumber \\ 
 &\qquad \times \Ex{}{e^{-\zeta \sum_{k=2}^{\ns} \infden\ltrp{\rndx_{k,\ell}; \rndy_{k,\ell}, \hat{h}_\ell}} \ggiven \rndb_{1,\ell} = j_1} 
    \label{eq:factorized:MGF}
\end{align}
where \eqref{eq:phi:LTE} follows from the law of total expectation and \eqref{eq:phi:conditional:independence} follows since $e^{-\zeta \infden\ltrp{\rndx_{1,\ell}; \rndy_{1,\ell},
\hat{h}_\ell}}$ and $e^{-\zeta \sum_{k=2}^{\ns} \infden\ltrp{\rndx_{k,\ell}; \rndy_{k,\ell}, \hat{h}_\ell}}$ are independent conditioned on $B_{1,\ell}$. The last factor in~\eqref{eq:factorized:MGF} can be written as
\begin{align}
    &\Ex{}{e^{-\zeta \sum_{k=2}^{\ns}
        \infden\ltrp{\rndx_{k,\ell}; \rndy_{k,\ell}, \hat{h}_\ell}} \ggiven
		\rndb_{1,\ell} = j_1}\notag\\ 
	&= \sum_{j_{\ns}=1}^{u^2}
        \Exop\Big[e^{-\zeta \sum_{k=2}^{\ns}
		\infden\ltrp{\rndx_{k,\ell}; \rndy_{k,\ell}, \hat{h}_\ell}} \Big | \rndb_{\ns,\ell} = j_{\ns}, \nonumber \\
  & \qquad \qquad \qquad \quad \rndb_{1,\ell} = j_1\Big]\Prob[B_{\ns} = j_{\ns}\mid B_1=j_1]\\
	&= \sum_{j_{\ns}=1}^{u^2}  \left[\matP_{\ell}(\zeta)^{\ns-1}\right]_{j_1j_{\ns}}.
    \label{eq:middle:factor}
\end{align}
Combining \eqref{eq:factorized:MGF} and \eqref{eq:middle:factor} yields
\begin{align}
    \varphi_{\ell}(\zeta) 
    &= \sum_{j_1=1}^{u^{2}}\sum_{j_{\ns}=1}^{u^2} \frac{\varphi_{j_1, \ell}(\zeta)}{u^2}  \left[\matP_{\ell}(\zeta)^{\ns-1}\right]_{j_1j_{\ns}}\\
    &= \boldsymbol{\nu}_\ell(\zeta)^{T}\matP_\ell(\zeta)^{\ns-1} \vecone_{u^2}, \label{eq:MGFVarphiL}
\end{align}
where
\begin{equation}\label{eq:boldnu}
\boldsymbol{\nu}_\ell(\zeta) = \left[\frac{1}{u^{2}} \varphi_{1,\ell}(\zeta), \ldots, \frac{1}{u^{2}}
    \varphi_{u^2,\ell}(\zeta)\right]^T.
\end{equation}

To summarize,~\eqref{eq:MGFVarphiL} provides an efficient way to evaluate the moment generating
function $\varphi_{\ell}(\zeta)$, provided that one can compute efficiently powers of the matrix
$\matP_{\ell}(\zeta)$.
Whether this is possible or not depends on the constellation that is used.
In the next section, we discuss in detail the practically relevant case of BPSK constellation.

\subsection{A Case Study: BPSK Constellation}
\label{Sec:BPSK_Mod}
We next consider BPSK constellation and use the mapping between $(X_{k-1,\ell}, X_{k,\ell})$ and
$B_{k,\ell}$ described in Table~\ref{tab:table1}.
It follows by symmetry that $\varphi_{1,\ell}(\zeta)=\varphi_{4,\ell}(\zeta)$ and that 
$\varphi_{2,\ell}(\zeta)=\varphi_{3,\ell}(\zeta)$.
As illustrated in Appendix \ref{Appen:NumEval}, these quantities and their derivatives, which are needed to numerically evaluate the saddlepoint expansion in
Theorem~\ref{thm:saddlepoint-generic}, can be computed efficiently using numerical
integration.

Using the symmetry just unveiled, we can express $\boldsymbol{\nu}_\ell(\zeta)$ and
$\matP_{\ell}(\zeta)$ as
\begin{equation}
     \boldsymbol{\nu}_{\ell}(\zeta) = \mat \frac{\varphi_{1,\ell}(\zeta)}{4} &  \frac{\varphi_{2,\ell}(\zeta)}{4} & \frac{\varphi_{2,\ell}(\zeta)}{4}  & \frac{\varphi_{1,\ell}(\zeta)}{4} \emat^{T},
\end{equation}
and
\begin{equation}
    \matP_{\ell}(\zeta) = \mat 
    \frac{\varphi_{1,\ell}(\zeta)}{2} & \frac{\varphi_{2,\ell}(\zeta)}{2} & 0                  & 0 \\
    0                  & 0                  & \frac{\varphi_{2,\ell}(\zeta)}{2} & \frac{\varphi_{1,\ell}(\zeta)}{2} \\
    \frac{\varphi_{1,\ell}(\zeta)}{2} & \frac{\varphi_{2,\ell}(\zeta)}{2} & 0                  & 0 \\
    0                  & 0                  & \frac{\varphi_{2,\ell}(\zeta)}{2} & \frac{\varphi_{1,\ell}(\zeta)}{2}
    \emat.
\end{equation}
To evaluate $\matP_{\ell}(\zeta)^{\ns-1}$ we write $\matP_{\ell}(\zeta)$ in terms of its eigenvalue
decomposition as $\matP_{\ell}(\zeta)=\matV_{\ell}\boldsymbol{\Lambda}_{\ell}\matV_{\ell}^{-1}$ where 
\begin{equation}
    \matV_{\ell} =  
    \mat 
    0 & -\frac{\varphi_{2,\ell}(\zeta)}{\varphi_{1,\ell}(\zeta)} & -1 & 1 \\ 
    0 & 1 & 1 & 1 \\ 
    -\frac{\varphi_{1,\ell}(\zeta)}{\varphi_{2,\ell}(\zeta)} & 0 & -1 & 1 \\ 
    1 & 0 & 1 & 1
    \emat 
\end{equation}
contains the eigenvectors of $\matP_{\ell}(\zeta)$ and 
\begin{equation}
    \boldsymbol{\Lambda}_{\ell} = \mat
    0 & 0 & 0 & 0 \\ 
    0 & 0 & 0 & 0 \\ 
    0 & 0 & \frac{\varphi_{1,\ell}(\zeta)-\varphi_{2,\ell}(\zeta)}{2} & 0 \\ 
    0 & 0 & 0 & \frac{\varphi_{1,\ell}(\zeta)+\varphi_{2,\ell}(\zeta)}{2}
    \emat,
\end{equation}
contains the corresponding eigenvalues. 
Then we use that 
$\matP_{\ell}(\zeta)^{\ns-1} = \matV_{\ell} \boldsymbol{\Lambda}_{\ell}^{\ns-1} \matV_{\ell}^{-1}$.
It then follows from~\eqref{eq:MGFVarphiL} that 
\begin{IEEEeqnarray}{rCl}
    \varphi_{\ell}(\zeta) = \frac{1}{2^{\ns}} \ltrp{\varphi_{1,\ell}(\zeta) +
    \varphi_{2,\ell}(\zeta)}^{\ns}.  
\end{IEEEeqnarray}

\section{Numerical Results and Discussion}
\label{sec:NumericalResults}
In this section, we report numerical experiments illustrating the performance of our synchronization method both in terms of normalized mean square error (NMSE) and achievable packet error probability. We also verify the accuracy of the saddlepoint approximation. 

Throughout, we shall consider the case of fully dependent delays; we shall also assume that a
BPSK constellation is used for both pilot and data transmission, and that m-sequences~\cite[Ch. 8]{proakisBook} are used as pilot sequences. We consider a Rayleigh-fading scenario, i.e., we assume that the $\rndh_{\ell}$, $\ell = 1, \ldots, \nb,$ are generated independently from a $\jpg(0,1)$ distribution. Since $N_0=1$, we can interpret the power constraint $\rho$ introduced in Section~\ref{sec:SystemModel} as the (average) SNR. We will analyze the performance of both the per-block and the joint synchronization algorithms proposed in Sections \ref{subsubsec:ind} and \ref{subsubsec:joint}, respectively. Note that per-block synchronization, when applied to case of fully dependent delays, achieves the same performance as if the delays across blocks were independent. 

To assess the performance of the estimators introduced in Section \ref{subsubsec:ind} and Section \ref{subsubsec:joint}, we compare their NMSE with the CRB, whose evaluation is detailed in Appendix \ref{Appen:CRB}. The delay NMSE is defined as $\Ex{}{(\rndd-\hat{\rndd})^2/ \tp^2}$ for the case of joint synchronization, and as $\Ex{}{(\rndd - \hat{\rndd}_{\ell})^2 / \tp^2}$ for the case of per-block synchronization. Note that this last quantity does not depend on $\ell$. The channel-gain NMSE is defined as $\Ex{}{\abs{\rndh_{\ell} - \hat{\rndh}_{\ell}}^2/\abs{\rndh_{\ell}}^2}$ for both algorithms. This quantity also does not depend on $\ell$.

\begin{figure}
\centering
   \begin{minipage}{\columnwidth}
  \centering
  \subfloat[Synchronization error]{\includegraphics[width=0.9\columnwidth,keepaspectratio]{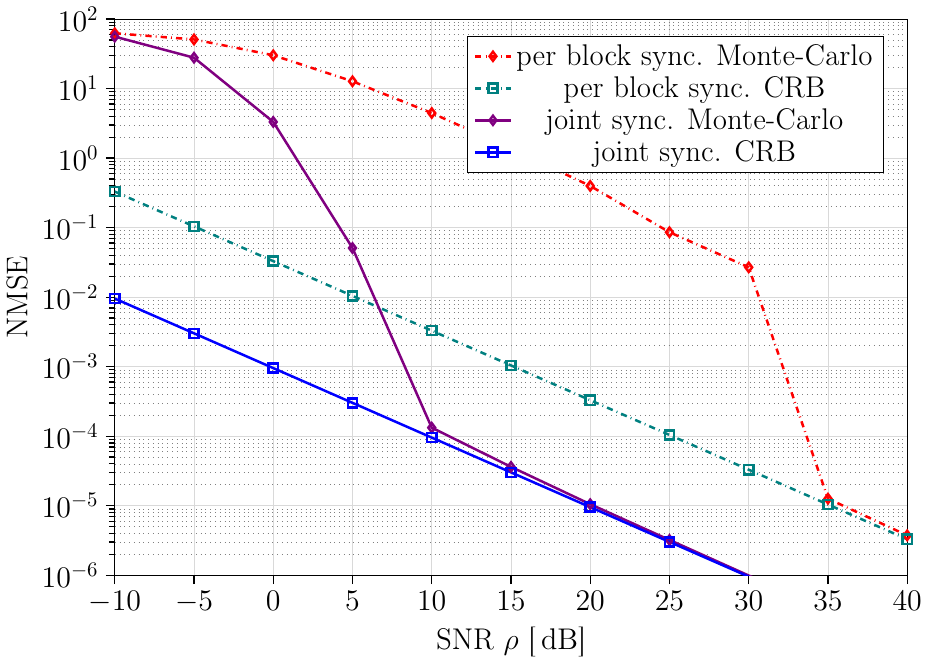}}
  \end{minipage}%
  
  \begin{minipage}{\columnwidth}
  \centering
  \subfloat[Channel estimation error]{\includegraphics[width=0.9\columnwidth,keepaspectratio]{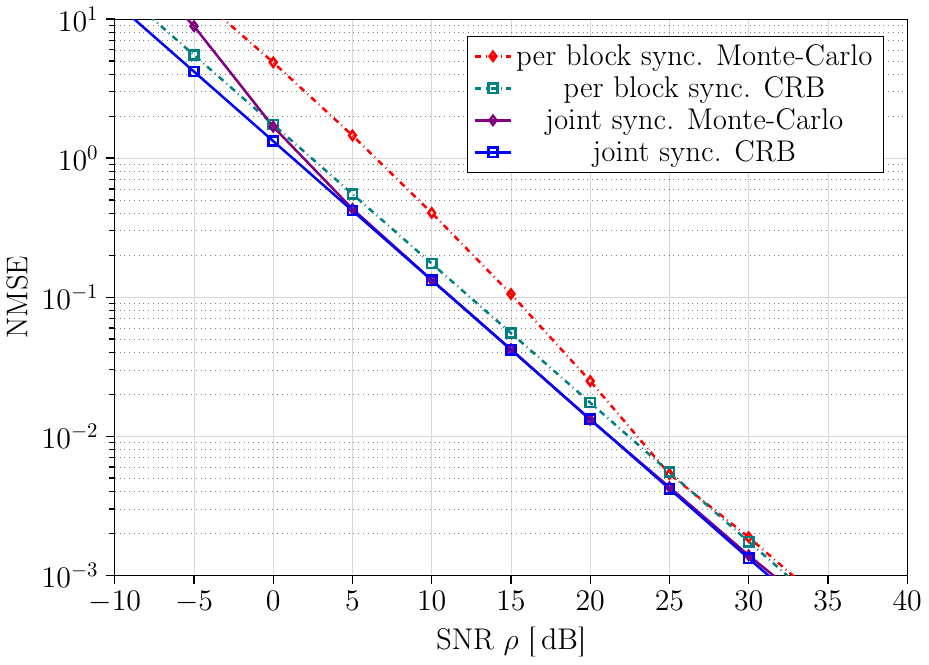}}
  \end{minipage}
  \caption{Comparison between NMSE of the estimators and CRB for $\mN=10$, $\np=7$, $\nb=4$.
  }
  \label{fig:CRB1}
\end{figure}

In Fig. \ref{fig:CRB1}, we depict the NMSE as a function of the SNR for $\mN = 10$, $\np=7$, and $\nb=4$. The curves are obtained by assuming no interference from the data symbol, i.e., by setting $\rvecc_{\ell} = \boldsymbol{0}_{\mM}$ in \eqref{eq:YpMMform}. Under this assumption, our estimators are unbiased and the NMSE approaches the CRB as the SNR increases, as shown in Fig. \ref{fig:CRB1}. As expected, joint synchronization significantly outperforms per-block synchronization (which, however, is the only option if the delays are independent across blocks). Furthermore, the convergence to the CRB is more rapid. 
%

\begin{figure}
    \centering
    \includegraphics[width=0.9\columnwidth]{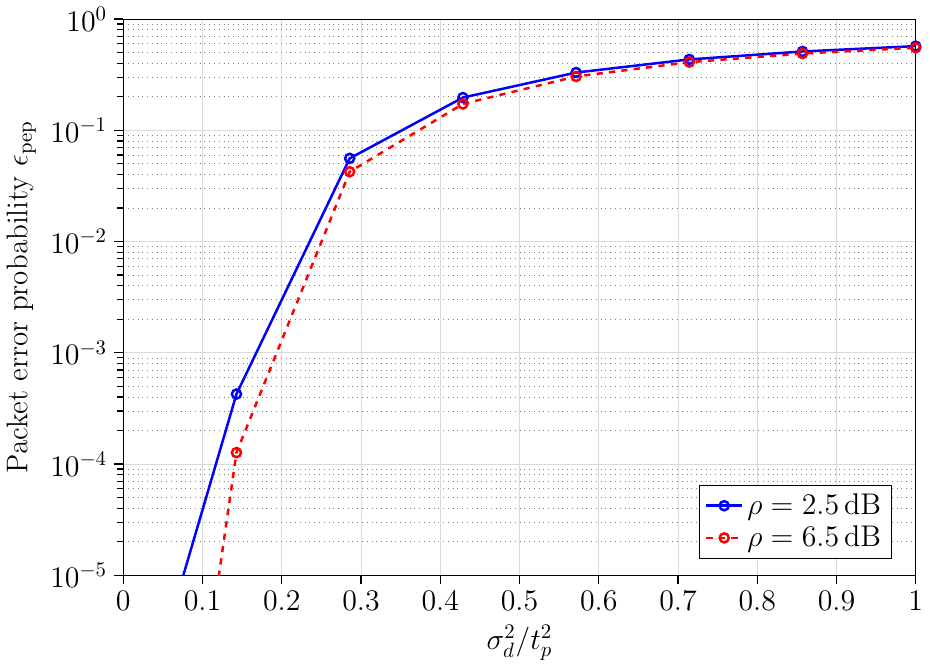}
    \caption{Achievable packet error probability evaluated using the RCUs bound as a function of the average synchronization error. Here $n=288$, $\nb =8$, $\rate = 0.104$ bit per channel use; $s$ is optimized.}
    \label{fig:d_err_effect}
\end{figure}

In Fig. \ref{fig:d_err_effect}, we illustrate the impact of synchronization errors on the packet error probability, evaluated by substituting the RCUs bound $\epsilon\sub{ub}$ for $\epsilon\sub{1}$ in \eqref{eqn:error-bound}. To do so, we let $\sigma_d^2$ be the mean squared error incurred when estimating $\rndd$. In this analysis (and only in this analysis), we assume the channel to be known at the receiver (i.e., $\hat{h}_\ell=h_\ell$ for all $\ell$), let $\hat{\rndd} \sim \mathcal{N}(\rndd,\sigma_d^2)$ (or $\hat{\rndd}_{\ell} \sim \normal(\rndd,\sigma_d^2)$ for the per-block synchronization algorithm), where we allow $\sigma_d^2$, which corresponds to the mean square error for delay estimation, to vary independently from any other system parameter, and report an upper bound on the $\epsilon\sub{pep}$ as a function of $\sigma_d^2/\tp^2$ for $\rho \in \{2.5, 6.5\}\dB$, $\nb\nc = 288$, $\nb = 8$, $\np =0$, and $\rate = 30/288 = 0.104$ bit per channel use.
These parameters reflect a URLLC scenario involving the transmission of
		      compact downlink control information~\cite{Ferrante2018,RAN_R1-1720997-Ericsson}.

The parameter $s$ of the RCUs bound is optimized. Note that, since we assumed that the $\{h_{\ell}\}$ are known to the receiver, once we fix $\sigma_d^2$, the packet error probability achieved using joint synchronization and per-block synchronization coincide.
We see from the figure that to achieve an error probability $\epsilon\sub{pep} < 10^{-4}$ for both values of $\rho$, it is enough that $\sigma_d^2/\tp^2$ is below $0.12$ and that the error probability deteriorates rapidly once this value is exceeded. 

\begin{figure}
    \centering
    \includegraphics[width=0.9\columnwidth]{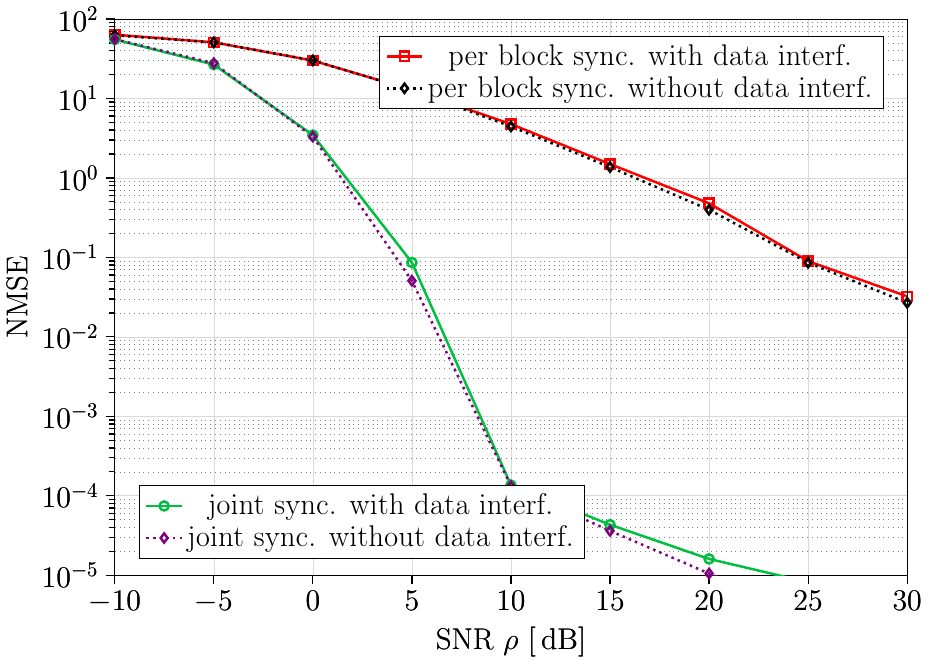}
    \caption{NMSE for the delay estimation for both joint and per-block synchronization for $\mN = 10$, $\nb=4$, $\np=7$.}
    \label{fig:h_d_error}
\end{figure}

We next assess the validity of the assumption that $\rvecc_{\ell} = \boldsymbol{0}_{\mM}$ by plotting in Fig. \ref{fig:h_d_error} the NMSE for the delay estimation as a function of the SNR for both the case of no data interference and data interference. In the figure, we assume that $N = 10$, $\nb = 4$, and $\np = 7$. As shown in the figure, the impact of data interference on the performance of our algorithm is minimal.

\begin{figure}[t]
    \centering
    \includegraphics[width=0.9\columnwidth]{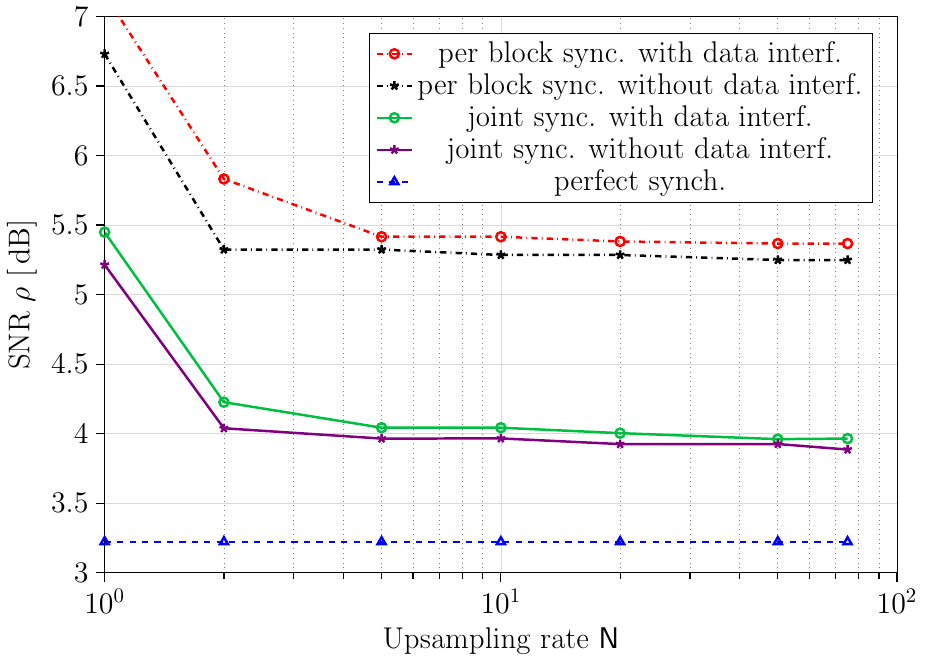}
    \caption{Upper bound on the SNR required to achieve $\epsilon_{\text{pep}} = 10^{-5}$ as a function of $\mN$. Here, $\nb\nc=288$, $\rate = 0.104$ bit per channel use and $\nb=8$; $\np$ and $s$ are optimized. } 
    \label{fig:Results_FigOverN}
\end{figure}

Next, we report in Fig. \ref{fig:Results_FigOverN} the SNR required to achieve a packet error rate $\epsilon = 10^{-5}$ for a transmission rate $\rate=0.104$ bit per channel use, $\nb=8$, and $\nb\nc = 288$ as a function of the upsampling rate $\mN$. The SNR values are obtained by performing an optimization over both $s$ and the number of pilot symbols $\np$. We see from the figure that a significant SNR reduction can be achieved by increasing $\mN$ from $1$ to $5$. However, a further increase in $\mN$ yields negligible performance improvements\footnote{The same trend is observed for the channel estimation and synchronization errors: the corresponding NMSE either does not improve much for $\mN>5$ or is small enough at $\mN=5$ such that further reduction does not significantly improve the error probability.}. In the figure, we also assess once more the impact of interference from the data symbols and conclude again that it is negligible.

\begin{figure}
    \centering
    \includegraphics[width=0.9\columnwidth]{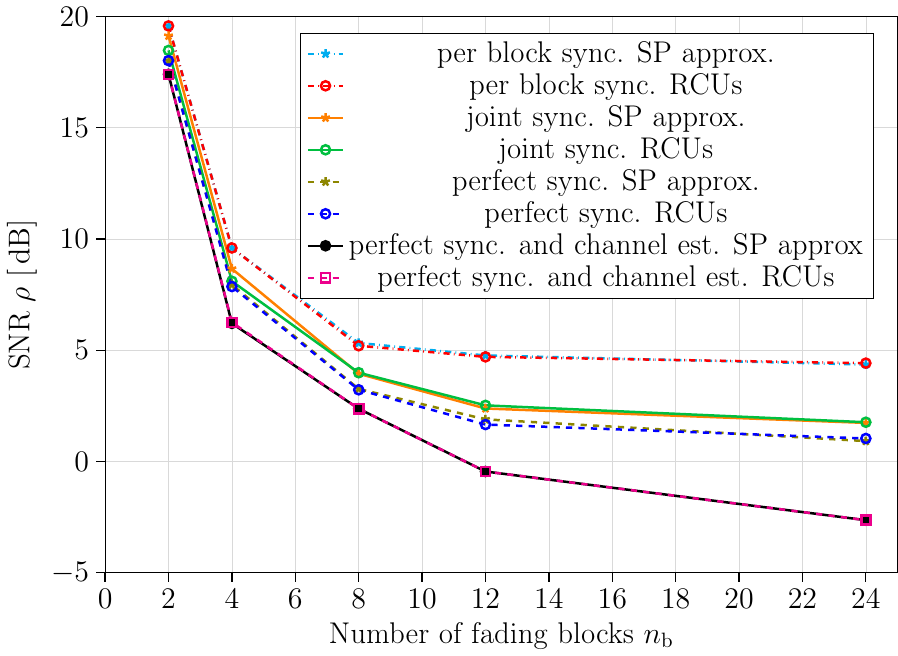}
    \caption{Upper bound on the SNR sufficient to achieve $\epsilon_{\text{pep}}=10^{-5}$ as a function of $\nb$. Here, $\nb\nc=288$, $\rate = 0.104$ bit per channel use, $\mN = 5$; $\np$ and $s$ are optimized.}
    \label{fig:Results_FigOvernb}
\end{figure}

In Fig. \ref{fig:Results_FigOvernb}, we analyze the performance achievable in the URLLC regime using the synchronization and channel estimation algorithms introduced in Section \ref{sec:RCUsSection}. Specifically, we show the SNR required to achieve an error probability of $10^{-5}$ for $\nb\nc=288$, $\mN =5$, and $\rate=0.104$ bit per channel, as a function of the number of fading blocks $\nb$ spanned by each codeword. We obtain each value of SNR by optimizing over both the number of pilot symbols $\np$ and the~$s$ parameter in the RCUs. 
We consider both per-block and joint synchronization, and depict for reference also the curve corresponding to perfect synchronization and channel estimation, and the one corresponding to perfect synchronization, but pilot-aided estimation of the channel gain.
For each scenario, we plot both the RCUs and its saddlepoint approximation. We observe that the saddlepoint approximation provides an accurate and numerically efficient approximation of the RCUs bound for the parameters considered in the figure. This also suggests that the conditions required for the saddlepoint approximation, given in \eqref{eq:fourth-moment-constraint} and \eqref{eq:second moment constraint}, hold in our setup.

We observe from the figure that per-block synchronization requires up to $3.5\dB$ higher SNR than joint synchronization to achieve the same error probability. We note that when $\nb \geq 4$, the SNR gap between joint synchronization and perfect synchronization but pilot-aided channel estimation is no larger than $0.6\dB$. This suggests that the pilot symbols needed to estimate the fading coefficients in the perfect synchronization, pilot-aided channel estimation case are sufficient to also estimate the delay when the joint synchronization algorithm is used. 
To strengthen this claim, we report the optimum number of $\np$ for joint, and perfect synchronization cases in Table \ref{tab:table_np}. Unsurprisingly, the optimum number of $\np$ for the joint and perfect synchronization case are the same for all values of $\nb$ considered in Fig. \ref{fig:Results_FigOvernb}. It is interesting to note that the gap between the perfect synchronization, pilot-aided channel-estimation curve and the curve corresponding to perfect synchronization and channel estimation\footnote{To perform a fair comparison between the two curves, the one corresponding to perfect synchronization and channel estimation is obtained by using the $\np$ values in Table \ref{tab:table_np}, although these pilot symbols are not used.} increases as a function of SNR. This is because, as $\nb$ is increased for a fixed $\nc\nb$, the number of symbols $\nc$ that can be transmitted in a block decreases. As a result, the optimal value of $\np$ decreases as well, as illustrated in Table \ref{tab:table_np}, which causes a deterioration in the accuracy of the channel estimate and an increase of the error probability, for a fixed SNR. 

\begin{table}[]
    \centering
    \caption{Optimal values of $\np$ in Fig. \ref{fig:Results_FigOvernb}.}
    \begin{tabular}{@{}ccc@{}}
        $\nb$& joint sync. $\np$ & perfect sync. $\np$ \\ \hline
         2  &  31 & 31\\ 
         4 &   15 & 15\\ 
         8 &   15 & 15\\ 
         12 &  7 & 7\\ 
         24 &  3 & 3
    \end{tabular}
    
    \label{tab:table_np}
\end{table}

\begin{figure}
    \centering
    \includegraphics[width=0.9\columnwidth]{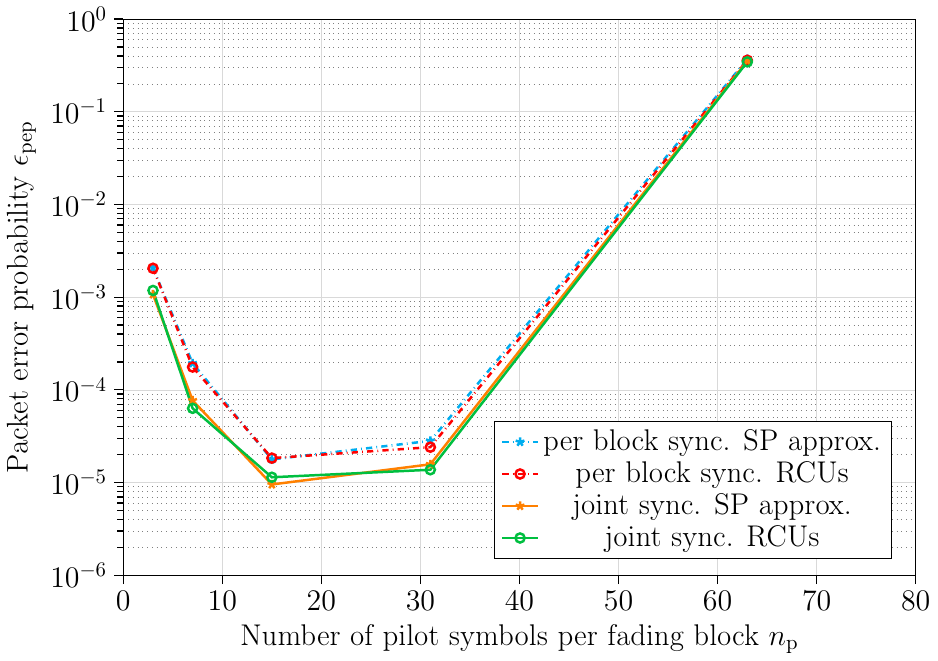}
    \caption{Upper bound on packet error probability $\epsilon_{\text{pep}}$ as a function of $\np$. Here, $\rho=8.45 \dB$, $\nb=4$, $\nb\nc=288$, $\rate = 0.104$ bit per channel use, $\mN = 5$; $s$ is optimized for each value of $\nb$. } 
    \label{fig:Results_FigOverNp}
\end{figure}

The optimal number of pilots for the per-block synchronization case turns out to coincide with that for the joint-synchronization case. The reason is as follows: while increasing $\np$ reduces estimation errors, it comes at the cost of a reduction of the number of data symbols $\ns$, which impacts the packet error probability. This is illustrated in Fig. \ref{fig:Results_FigOverNp}, where we report our upper bound on the packet error probability $\epsilon\sub{pep}$ as a function of $\np$ for $\rate = 0.104$ bit per channel use, $\nb = 4$, $\nc\nb=288$, $\mN = 5$, and $\rho=8.45\dB$ for both per-block and joint synchronization. We see that the value of $\np$ that minimizes our upper bound $\epsilon\sub{pep}$ coincides with per-block and joint synchronization. However, per-block synchronization yields a higher error probability for a fixed SNR. This implies that, as shown in Fig. \ref{fig:Results_FigOvernb}, per-block synchronization requires a higher SNR than joint-synchronization, to achieve a given target error probability.
The results in Fig.~\ref{fig:Results_FigOverNp} illustrate also that the saddlepoint approximation provides an accurate approximation for all $\np$ values considered in the figure. This is of utmost importance because the parameter $\np$ may not be optimized specifically to minimize the packet error probability in every system. Therefore, the ability to utilize the approximations, irrespective of the chosen value for $\np$, holds significant relevance.
%

\section{Conclusions}
\label{sec:conclusion}
We have presented an efficient method to evaluate an upper bound on the error probability achievable over memoryless block-fading channels, with pilot-assisted transmission for channel estimation and timing synchronization. The method is based on a novel saddlepoint approximation, which accounts for the dependence across certain random variables arising in the presence of synchronization errors. 

Numerical experiments conducted using two synchronization and channel-estimation algorithms show that the proposed saddlepoint approximation can be safely used to benchmark URLLC systems. We show how to use our approximation to determine the synchronization level required to achieve the low error probabilities demanded in URLLC applications. Moreover, our numerical results reveal that, when the delays are fully dependent across fading blocks and synchronization is performed jointly over the blocks, the pilot symbols needed for channel estimate are sufficient to acquire synchronization. However, if we use per-block synchronization (which is unavoidable if delays are independent across blocks), the same number of pilot symbols is not enough to achieve a sufficiently good synchronization. As a result, for the case when the delays are independent, synchronization becomes the bottleneck for the system performance, and yields an SNR penalty that can be quantified using our bounds. A generalization of the analysis presented in this paper to the case of multiple-antenna systems is an interesting topic for a future work.

\appendices
\section{Numerical Evaluation of \eqref{eq:varphi0j} and its first two derivatives}
\label{Appen:NumEval}
We consider BPSK constellation and use the mapping described in Table \ref{tab:table1}. For this case, \eqref{eq:InfoDens} reduces to
\begin{IEEEeqnarray}{rCl}
    \label{eq:InfoDensBPSK}
    \infden(x;y,\hat{h}) &=& -s\abs{y-\hat{h}x}^2 \nonumber \\ 
    && - \log \frac{e^{-s\abs{y - \hat{h} x^{(1)}}^2} + e^{-s\abs{y  - \hat{h} x^{(2)}}^2}}{2}.
\end{IEEEeqnarray}

For a given $\rndb_{k,\ell} = j$, we have that $X_{k-1,\ell}=x_{j1}$ and $X_{k,\ell}=x_{j2}$. Hence, 
conditioned on $\Delta_\ell=\delta_\ell$, $H_\ell=h_\ell$, $\hat H_\ell=\hat h_\ell$, we have
\begin{align}
    \rndy_{k,\ell} &= h_\ell(\delta x_{j2} + (1-\delta) x_{j1}) + Z_{k,\ell} \\
                   &= \alpha_j + Z_{k,\ell}, 
\end{align}
where $\alpha_j = h_\ell(\delta x_{j2} + (1-\delta) x_{j1})$. 
Hence, 
\begin{align}
    \label{eq:InfoDensBPSK2}
    &\infden(X_{k,\ell}; Y_{k,\ell}, \hat{h}) = -s\abs{\alpha_j + Z_{k,\ell}-\hat{h}x}^2 \nonumber \\
     &- \log \frac{1}{2}\ltrp{e^{-s\abs{\alpha_j + Z_{k,\ell} - \hat{h} x^{(1)}}^2} 
    + e^{-s\abs{\alpha_j + Z_{k,\ell}  - \hat{h} x^{(2)}}^2}},
\end{align}
and
\begin{equation}
    e^{-\zeta\infden(X_{k,\ell}; Y_{k,\ell}, \hat{h})} = e^{\zeta f_1(Z_{k,\ell})}f_2(Z_{k,\ell})^\zeta,
\end{equation}
where
\begin{align}
    f_1(Z_{k,\ell}) 
    &= s \abs{\alpha_j + Z_{k,\ell} - \hat{h}_{\ell}x_{j2} }^2,\\ 
    f_2(Z_{k,\ell}) 
    &=\frac{1}{2} \ltrp{e^{-s\abs{\alpha_j + Z_{k,\ell} - \hat{h}_{\ell} x^{(1)}}^2} + e^{-s\abs{\alpha_j + Z_{k,\ell}  - \hat{h}_{\ell} x^{(2)}}^2} } .
\end{align}
The $n$th derivative of $\exp(-\zeta\infden(X_{k,\ell}; Y_{k,\ell}, \hat{h}))$ is 
\begin{align}
    &\frac{d^n}{d\zeta^n} e^{-\zeta\infden(X_{k,\ell}; Y_{k,\ell}, \hat{h})} 
     \nonumber \\
    & \quad = e^{\zeta f_1(Z_{k,\ell})}f_2(Z_{k,\ell})^\zeta  (f_1(Z_{k,\ell}) + \log f_2(Z_{k,\ell}))^n.
\end{align}
Hence,
\begin{align}
    \varphi_{j,\ell}(\zeta) &= \Exop\left[{e^{\zeta f_1(Z_{k,\ell})}f_2(Z_{k,\ell})^\zeta}\right] \label{eq:ES:phiDiv0}\\
    \varphi_{j,\ell}'(\zeta) &= \Exop\left[{e^{\zeta f_1(Z_{k,\ell})}f_2(Z_{k,\ell})^\zeta(f_1(Z_{k,\ell}) + \log f_2(Z_{k,\ell}))}\right]\label{eq:ES:phiDiv1} \\
    \varphi_{j,\ell}''(\zeta)
    &= \Exop\left[{e^{\zeta f_1(Z_{k,\ell})}f_2(Z_{k,\ell})^\zeta(f_1(Z_{k,\ell}) + \log f_2(Z_{k,\ell}))^2}\right]\label{eq:ES:phiDiv2}.
\end{align}
Since $\rndz_{k,\ell} \sim \jpg(0,1)$, the expectations in \eqref{eq:ES:phiDiv0}, \eqref{eq:ES:phiDiv1}, and \eqref{eq:ES:phiDiv2} can be expressed as two-dimensional real integrals, which can be evaluated efficiently via numerical integration methods.

\section{Evaluation of CRB}
\label{Appen:CRB}
For the purpose of computing the CRB in the asychronous case, we consider the problem of estimating the parameter vector $\boldsymbol{\theta} = [\re\{\vech^{T}\}, \im\{\vech^{T}\}, \vecd^T]^{T}\in\mathbb{R}^{3\nb}$ where $\vecd =~ \ltrsqr{d_1,\ldots, d_{\nb}}^T$.
We view the parameter vector as unknown and deterministic. Ignoring the the data interference part in \eqref{eq:YpMMform}, the observed data vector is 
\begin{align}
    \rvecy^{(p)} = 
    \ltrsqr{h_1 \vecv(d_1)^T, \ldots, h_{\nb} \vecv(d_{\nb})}^T
    +\ltrsqr{\rvecz_1^T, \ldots, \rvecz_{\nb}^T}^T  \label{eq:CRB:data:vector}
\end{align}
where $\mM$ is defined in~\eqref{eq:Mdef},
\begin{equation}
    \vecv(d_\ell) =
    \left(1 - \frac{e_\ell}{\ts}\right)\vecx^{(p)}_{\mN}(q_\ell) + \frac{e_\ell}{\ts}\vecx^{(p)}_{\mN}(q_\ell + 1)
     \label{eq:CRB:v:vector}
\end{equation}
and $q_\ell = \lfloor d_\ell/\ts \rfloor$ and $e_\ell = d_\ell - q_\ell\ts$.
Hence, $\rvecy^{(p)}$ is a complex Gaussian random vector with covariance matrix $\matI_{\nb\mM}$ and mean vector
\begin{equation}
    \boldsymbol{\mu}(\boldsymbol{\theta}) = 
    \ltrsqr{h_1 \vecv(d_1)^T, \ldots, h_{\nb} \vecv(d_{\nb})^T}^T.\label{eq:CRB:mu:vector}
\end{equation}
The Fisher information matrix $\matJ(\boldsymbol{\theta})$ for this problem is given by \cite[Ch. 15.7]{Kay_1993}
\begin{equation}
\ltrsqr{\matJ(\boldsymbol{\theta})}_{mn} 
= 2\re\ltrcurley{ \frac{\partial \boldsymbol{\mu}(\boldsymbol{\theta})}{\partial \theta_m}^H 
\frac{\partial \boldsymbol{\mu}(\boldsymbol{\theta})}{\partial \theta_n}},
\end{equation}
where $m,n \in \{1,\dots, 3\nb\}$, $\theta_m$ is the $m$th element of $\boldsymbol{\theta}$, and the derivatives are evaluated at the true parameter values. Hence, we need to compute the derivatives of $\boldsymbol{\mu}(\boldsymbol{\theta})$ with respect to the real and imaginary parts of $h_1, \ldots, h_{\nb}$, which is trivial, and with respect to $d_1,\ldots, d_{\nb}$, which requires a bit of care. We note that if $e_\ell\neq 0$, then
\begin{equation}
    \frac{\partial}{\partial d_\ell}h_\ell\vecv(d_\ell) =
    \frac{h_\ell}{\ts}(\vecx^{(p)}_{\mN}(q_\ell + 1) -\vecx^{(p)}_{\mN}(q_\ell)).
\end{equation}
However, if $e_\ell= 0$, then $\boldsymbol{\mu}(\boldsymbol{\theta})$ may not be differentiable with respect to $d_\ell$, and the CRB cannot be computed for this case. We will therefore from now on assume\footnote{This is not a strong assumption, since $E_\ell = 0$ occurs with probability zero.} that $e_\ell\neq 0$ for $\ell \in\{1, 2, \ldots, \nb\}$. It then follows from \eqref{eq:CRB:v:vector} and \eqref{eq:CRB:mu:vector} that, for $\ell \in\{1, 2, \ldots, \nb\}$, 
\begin{align}
    \frac{\partial \boldsymbol{\mu}(\boldsymbol{\theta})}{\partial \theta_{\ell}}
    &=  \ltrsqr{\mathbf{0}_{\mM (\ell-1)}^{T}, \vecv(d_{\ell})^{T}, \mathbf{0}_{\mM(\nb-\ell)}^{T} }^{T}\\
    \frac{\partial \boldsymbol{\mu}(\boldsymbol{\theta})}{\partial \theta_{\nb+\ell}}
    &=  \ltrsqr{\mathbf{0}_{\mM (\ell-1)}^{T}, j\vecv(d_{\ell})^{T}, \mathbf{0}_{\mM(\nb-\ell)}^{T} }^{T}\\
    \frac{\partial \boldsymbol{\mu}(\boldsymbol{\theta})}{\partial \theta_{2\nb+\ell}}
    &=  \frac{h_\ell}{\ts}\ltrsqr{\mathbf{0}_{\mM (\ell-1)}^{T}, \vecx^{(p)}_{\mN}(q_\ell+1) - \vecx^{(p)}_{\mN}(q_\ell), \mathbf{0}_{\mM(\nb-\ell)}^{T} }^{T} \hspace{-0.2em}. \label{eq:delay:derivative}
\end{align}

Using the Fisher information matrix $\matJ(\boldsymbol{\theta})$, we can bound the estimation error variances for unbiased estimators as
\begin{align}
    \Ex{}{\left(\re\{\hat{{\rndh}_\ell}\}-\re\{{h_\ell}\}\right)^2 } &\geq \ltrsqr{\matJ(\boldsymbol{\theta})^{-1}}_{\ell\ell} \label{eq:CRBeq1}\\ 
    \Ex{}{\left(\im\{{\hat{\rndh}_\ell}\}-\im\{{h_\ell}\}\right)^2 } &\geq \ltrsqr{\matJ(\boldsymbol{\theta})^{-1}}_{(\nb+\ell)(\nb+\ell)} \label{eq:CRBeq2} \\ 
    \Ex{}{\left(\hat{D}_\ell-d_\ell\right)^2 } &\geq \ltrsqr{\matJ(\boldsymbol{\theta})^{-1}}_{(2\nb+\ell)(2\nb + \ell)} \label{eq:CRBeq3}
\end{align}
where $\ell \in \{1 ,\dots, \nb\}$. We may obtain the CRB for the channel estimation as
\begin{align}
    &\Ex{}{\abs{\hat{\rndh}_\ell - h_\ell}^2} \nonumber \\
    &= \Ex{}{(\re\{\hat{\rndh}_\ell\} - \re\{h_\ell\})^2 
    + (\im\{\hat{\rndh}_\ell\} - \im\{h_\ell\})^2}
     \label{eq:CRB_unbiased}\\
    & \geq \ltrsqr{\matJ(\boldsymbol{\theta})^{-1}}_{\ell\ell} + \ltrsqr{\matJ(\boldsymbol{\theta})^{-1}}_{(\ell+\nb)(\ell+\nb)}. \label{eq:CRB_unbiased2}
\end{align}

For the case of fully dependent delays, there is only one delay parameter $d=d_1$. However, the above derivation still holds if we redefine the parameter vector as $\boldsymbol{\theta} =~ [\re\{\vech^{T}\}, \im\{\vech^{T}\}, d]^{T}\in~\mathbb{R}^{2\nb+1}$, set $d_\ell=d_1$ for $\ell = 2, \ldots, \nb$, and restrict \eqref{eq:delay:derivative} and \eqref{eq:CRBeq3} to hold only for $\ell=1$. Of course, the dimension of the Fisher information matrix is also reduced to be $(2\nb + 1)\times(2\nb + 1)$.


\begin{thebibliography}{10}
\providecommand{\url}[1]{#1}
\csname url@samestyle\endcsname
\providecommand{\newblock}{\relax}
\providecommand{\bibinfo}[2]{#2}
\providecommand{\BIBentrySTDinterwordspacing}{\spaceskip=0pt\relax}
\providecommand{\BIBentryALTinterwordstretchfactor}{4}
\providecommand{\BIBentryALTinterwordspacing}{\spaceskip=\fontdimen2\font plus
\BIBentryALTinterwordstretchfactor\fontdimen3\font minus
  \fontdimen4\font\relax}
\providecommand{\BIBforeignlanguage}[2]{{%
\expandafter\ifx\csname l@#1\endcsname\relax
\typeout{** WARNING: IEEEtran.bst: No hyphenation pattern has been}%
\typeout{** loaded for the language `#1'. Using the pattern for}%
\typeout{** the default language instead.}%
\else
\language=\csname l@#1\endcsname
\fi
#2}}
\providecommand{\BIBdecl}{\relax}
\BIBdecl

\bibitem{Kislal2024_ICCPaper}
A.~O. Kislal, M.~Rajiv, G.~Durisi, E.~G. {Ström}, and U.~Mitra,
  ``Pilot-assisted {URLLC} links: Impact of synchronization error,'' in
  \emph{Proc. IEEE Int. Conf. Commun. (ICC)}, Denver CO, USA, Jun. 2024, pp.
  617--622.

\bibitem{Kolovou2021}
G.~Kolovou, S.~Oteafy, and P.~Chatzimisios, ``A remote surgery use case for the
  {IEEE} p1918.1 tactile internet standard,'' in \emph{Proc. IEEE Int. Conf.
  Commun. (ICC)}, Montreal, QC, Canada, Aug. 2021, pp. 1--6.

\bibitem{Liou2020}
E.-C. Liou and S.-C. Cheng, ``A {QoS} benchmark system for telemedicine
  communication over {5G} {uRLLC} and {mMTC} scenarios,'' in \emph{Proc. IEEE
  2nd Eurasia Conf. Biomed. Eng., Healthcare, Sustainability (ECBIOS)}, Tainan,
  Taiwan, Sep. 2020, pp. 24--26.

\bibitem{Peng2023}
Q.~Peng, H.~Ren, C.~Pan, N.~Liu, and M.~Elkashlan, ``Resource allocation for
  uplink cell-free massive {MIMO} enabled {URLLC} in a smart factory,''
  \emph{IEEE Trans. Commun.}, vol.~71, no.~1, pp. 553--568, Nov. 2023.

\bibitem{3GPP_URLLC}
\BIBentryALTinterwordspacing
``Study on physical layer enhancements for {NR} ultra-reliable and low latency
  case ({URLLC}) (release 16),'' {3GPP}, Tech. Rep., Mar. 2019. [Online].
  Available:
  \url{https://portal.3gpp.org/desktopmodules/Specifications/SpecificationDetails.aspx?specificationId=3498}
\BIBentrySTDinterwordspacing

\bibitem{Tataria2021}
H.~Tataria, M.~Shafi, A.~F. Molisch, M.~Dohler, H.~Sjöland, and F.~Tufvesson,
  ``{6G} wireless systems: Vision, requirements, challenges, insights, and
  opportunities,'' \emph{Proc. {IEEE}}, vol. 109, no.~7, pp. 1166--1199, Mar.
  2021.

\bibitem{Chentao2023}
C.~Yue, V.~Miloslavskaya, M.~Shirvanimoghaddam, B.~Vucetic, and Y.~Li,
  ``Efficient decoders for short block length codes in {6G} {URLLC},''
  \emph{IEEE Commun. Mag.}, vol.~61, no.~4, pp. 84--90, 2023.

\bibitem{Coskun2019}
M.~C. Coşkun, G.~Durisi, T.~Jerkovits, G.~Liva, W.~Ryan, B.~Stein, and
  F.~Steiner, ``Efficient error-correcting codes in the short blocklength
  regime,'' \emph{Physical Communication}, vol.~34, pp. 66--79, 2019.

\bibitem{durisi16-09a}
G.~Durisi, T.~Koch, and P.~Popovski, ``Towards massive, ultra-reliable, and
  low-latency wireless communication with short packets,'' \emph{Proc. {IEEE}},
  vol. 104, no.~9, pp. 1711--1726, Sep. 2016.

\bibitem{Polyanskiy2010}
Y.~{Polyanskiy}, H.~V. {Poor}, and S.~{Verd\'u}, ``Channel coding rate in the
  finite blocklength regime,'' \emph{{IEEE} Trans. Inf. Theory}, vol.~56,
  no.~5, pp. 2307--2359, May 2010.

\bibitem{Anand2018}
A.~Anand and G.~de~Veciana, ``Resource allocation and {HARQ} optimization for
  {URLLC} traffic in {5G} wireless networks,'' \emph{{IEEE} J. Sel. Areas
  Commun.}, vol.~36, no.~11, pp. 2411--2421, Oct. 2018.

\bibitem{Ghanem2022}
W.~R. Ghanem, V.~Jamali, M.~Schellmann, H.~Cao, J.~Eichinger, and R.~Schober,
  ``Codebook based two-time scale resource allocation design for {IRS}-assisted
  {eMBB-URLLC} systems,'' in \emph{IEEE Globecom Workshops (GC Wkshps)}, Rio de
  Janeiro, Brazil, Dec. 2022, pp. 419--425.

\bibitem{Darabi2022}
M.~Darabi, V.~Jamali, L.~Lampe, and R.~Schober, ``Hybrid puncturing and
  superposition scheme for joint scheduling of {URLLC} and {eMBB} traffic,''
  \emph{IEEE Commun. Lett.}, vol.~26, no.~5, pp. 1081--1085, Feb. 2022.

\bibitem{Chen2023}
Q.~Chen, J.~Wu, J.~Wang, and H.~Jiang, ``Coexistence of {URLLC} and {eMBB}
  services in {MIMO-NOMA} systems,'' \emph{IEEE Trans. Veh. Technol.}, vol.~72,
  no.~1, pp. 839--851, Sep. 2023.

\bibitem{jensen95-a}
J.~L. Jensen, \emph{Saddlepoint {A}pproximations}.\hskip 1em plus 0.5em minus
  0.4em\relax Oxford, U.K.: Oxford Univ. Press, 1995.

\bibitem{Martinez2011}
A.~{Martinez} and A.~{Guill\'en i F{\`a}bregas}, ``Saddlepoint approximation of
  random-coding bounds,'' in \emph{Proc. Inf. Theory Appl. Workshop}, San
  Diego, CA, USA, Feb. 2011, pp. 257--262.

\bibitem{Ostman2019}
J.~{{\"O}stman}, G.~{Durisi}, E.~G. {Str{\"o}m}, M.~C. {Co{\c s}kun}, and
  G.~{Liva}, ``Short packets over block-memoryless fading channels:
  Pilot-assisted or noncoherent transmission?'' \emph{{IEEE} Trans. Commun.},
  vol.~67, no.~2, pp. 1521--1536, Feb. 2019.

\bibitem{Lapidoth2002}
A.~{Lapidoth} and S.~{Shamai (Shitz)}, ``Fading channels: how perfect need
  "perfect side information" be?'' \emph{{IEEE} Trans. Inf. Theory}, vol.~48,
  no.~5, pp. 1118--1134, May 2002.

\bibitem{Kislal2023}
A.~O. Kislal, A.~Lancho, G.~Durisi, and E.~G. Ström, ``Efficient evaluation of
  the error probability for pilot-assisted {URLLC} with massive {MIMO},''
  \emph{IEEE J. Sel. Areas Commun.}, vol.~41, no.~7, pp. 1969--1981, May 2023.

\bibitem{Feller1971}
W.~Feller, \emph{An {I}ntroduction to {P}robability {T}heory and {I}ts
  {A}pplications}.\hskip 1em plus 0.5em minus 0.4em\relax New York, NY, USA:
  Wiley, 1971, vol.~2.

\bibitem{Lancho2020}
A.~{Lancho}, J.~{{\"O}stman}, G.~{Durisi}, T.~{Koch}, and G.~{Vazquez-Vilar},
  ``Saddlepoint approximations for short-packet wireless communications,''
  \emph{{IEEE} Trans. Wireless Commun.}, vol.~19, no.~7, pp. 4831--4846, Jul.
  2020.

\bibitem{Ferrante2018}
G.~C. {Ferrante}, J.~{{\"O}stman}, G.~{Durisi}, and K.~{Kittichokechai},
  ``Pilot-assisted short-packet transmission over multiantenna fading channels:
  A {5G} case study,'' in \emph{Proc. Conf. on Inf. Sci. and Sys. (CISS)},
  Princeton, NJ, U.S.A., Mar. 2018, pp. 1--6.

\bibitem{ostman2020}
J.~{{\"O}stman}, A.~Lancho, G.~Durisi, and L.~Sanguinetti, ``{URLLC} with
  massive {MIMO}: Analysis and design at finite blocklength,'' \emph{{IEEE}
  Trans. Wireless Commun.}, vol.~20, no.~10, pp. 6387--6401, Oct. 2021.

\bibitem{Yury2013}
Y.~Polyanskiy, ``Asynchronous communication: Exact synchronization,
  universality, and dispersion,'' \emph{IEEE Trans. Inf. Theory}, vol.~59,
  no.~3, pp. 1256--1270, 2013.

\bibitem{Scarlett2014}
J.~{Scarlett}, A.~{Martinez}, and A.~{Guill{\'e}n i F{\`a}bregas}, ``Mismatched
  decoding: Error exponents, second-order rates and saddlepoint
  approximations,'' \emph{{IEEE} Trans. Inf. Theory}, vol.~60, no.~5, pp.
  2647--2666, May 2014.

\bibitem{proakisBook}
J.~G. Proakis, \emph{Digital {C}ommunications}, 4th~ed.\hskip 1em plus 0.5em
  minus 0.4em\relax Boston, MA, USA: McGraw-Hill, 2008.

\bibitem{RAN_R1-1720997-Ericsson}
\BIBentryALTinterwordspacing
``R1-1720997: On {PDCCH} for ultra-reliable transmission,'' 3GPP, Tech. Rep.,
  Nov. 2017. [Online]. Available:
  \url{https://www.3gpp.org//ftp/TSG_RAN/WG1_RL1/TSGR1_91//docs/}
\BIBentrySTDinterwordspacing

\bibitem{Kay_1993}
S.~M. Kay, \emph{Fundamentals of Statistical Signal Processing, Volume I:
  Estimation Theory}.\hskip 1em plus 0.5em minus 0.4em\relax Englewood Cliffs:
  Prentice-Hall, 1993.

\end{thebibliography}
\end{document}